\documentclass[preprint,12pt,authoryear]{elsarticle}
\usepackage{amsmath,amssymb}
\usepackage{hyperref}

\usepackage{graphicx,caption,subcaption}
\usepackage{natbib}
\usepackage{multirow}
\usepackage{lineno}
\usepackage{aas_macros}
\usepackage{enumitem}
\modulolinenumbers[5]

\journal{Journal of \LaTeX\ Templates}

\biboptions{authoryear}

\journal{Advances in Space Research}

\newcommand{\disp}{\displaystyle}
\newcommand{\mbf}[1]{\mathbf{#1}}

\newcommand{\ol}[1]{\overline{#1}}

\newcommand{\dif}{\mathrm{d}}

\begin{document}

\begin{frontmatter}

\title{Topological and statistical properties of nonlinear force-free fields.}

\author[iia]{A.\ Mangalam\fnref{eid1} \corref{cor}}
\author[uso]{A.\ Prasad\fnref{eid2}}
\address[iia]{Indian Institute of Astrophysics, Sarjapur Road, Koramangala, Bangalore 560034, India}
\fntext[eid1]{mangalam@iiap.res.in}
\address[uso]{Udaipur Solar Observatory, Physical Research Laboratory, Dewali, Bari Road, Udaipur 313 001, India}
\fntext[eid2]{avijeet@prl.res.in}
\cortext[cor]{Corresponding author}
\begin{abstract}
We use our semi-analytic  solution of the nonlinear force-free field equation to construct three-dimensional magnetic fields that are applicable to the solar corona and study their statistical properties for estimating the degree of braiding exhibited by these fields. We present a new formula for calculating the winding number and compare it with the formula for the crossing number. The comparison is shown for a toy model of two helices and for realistic cases of nonlinear force-free fields; conceptually the formulae are nearly the same but the resulting distributions calculated for a given topology can be different. We also calculate linkages, which are useful topological quantities that are independent measures of the contribution of magnetic braiding to the total free energy and relative helicity of the field. Finally, we derive new analytical bounds for the free energy and relative helicity for the field configurations in terms of the linking number. These bounds will be of utility in estimating the braided energy available
for nano-flares or for eruptions.
 
\end{abstract}


\end{frontmatter}


\section{Introduction}
\label{s:intro}
The temperature of the solar corona is known to be around million degrees for decades \citep{1934ZA......8..124G, 1943ZA.....22...30E}. The average density of plasma in the corona is very low $\sim 10^8$ cm$^{-3}$ \citep{2004psci.book.....A}. The energy input required to compensate for the radiative and conductive losses and still maintain a million degree hot corona is estimated to be $10^7$ ergs cm$^2$s$^{-1}$ for active regions and $3\times 10^5$ ergs cm$^2$s$^{-1}$ for the quiet regions \citep{1977ARA&A..15..363W,2006SoPh..234...41K}. The physical processes that result in the heating of the corona are not well understood, though it is believed that a key role in this is played by the magnetic fields \citep{2000ssma.book.....S,2009soco.book.....G, 2015JPlPh..81d3904B}. The coronal heating theories can be broadly divided into two categories: direct current (DC) heating models, which are based on dissipation of magnetic stresses,  and alternating current (AC) heating models which are based dissipation of waves \citep{1985SoPh..100..289I,1997ApJ...490..442M,2000ApJ...530..999M,2006SoPh..234...41K}. In AC heating models, it is assumed that the photospheric motion changes on a time scale faster than what the coronal loop can adjust to (e.g., by damping and dissipation of Alfv{\'e}n waves), whereas in the DC heating models, it is assumed that the random photospheric motions displace the footpoints of the coronal magnetic field lines on time scales much longer than the Alfv{\'e}n transit time along a coronal loop, so that the loop can adjust to the changing boundary condition in a quasi-static way. Both AC and DC models involve photospheric footpoint motions which arise from the interactions of the convective plasma flows with the magnetic flux elements \citep{2014ApJ...787...87V}.

In the case of the DC heating models, the random rotations of the footpoints lead to twisting of the magnetic flux elements, while the random walks of these footpoints lead to their braiding \citep{1979cmft.book.....P,2009ApJ...705..347B}. In order to resist the increase in complexity, the coronal magnetic field in the corona tries to adjust its topology through continuous deformations. According to Parker's magnetostatic theorem \citep{1972ApJ...174..499P,1988ApJ...330..474P,1994ISAA....1.....P}, astrophysical plasmas with high magnetic Reynolds number and a complex magnetic topology favor spontaneous generation of current sheets (resulting from sharp gradients in the magnetic field) which leads to recurrent magnetic reconnections \citep{2016ApJ...830...80K}.  \citet{1972ApJ...174..499P,1983ApJ...264..642P}, \citet{1993PhRvL..70..705B} and \citet{2009ApJ...705..347B} then proposed a model which involves heating of the solar corona through nanoflares due to reconnection of braided magnetic flux elements. He further estimated the heating rate in the corona arising from the dynamical dissipation of the braided magnetic fields to be of the order of $10^7$ ergs cm$^{-2}$ s$^{-1}$ \citep{1983ApJ...264..642P} and argued it to be the principal source of heating in the active corona. The magnetic braiding can be characterized by defining a `crossing number' which can be related to the free energy of the field \citep{1993PhRvL..70..705B}. For continuous fields without distinct flux tube structures, some number $N$ of individual field lines can be chosen within a loop, and the braiding between these lines can be quantified. \citet{2009ApJ...696.1339W} presented such a semi-analytic force-free model of a pigtail braid  where three magnetic field lines crossed each other six times.

However, in these studies, simple analytic configurations of magnetic fields were considered that lacked the natural complexity often observed in active regions of the Sun. Model configurations of the coronal magnetic field that are morphologically similar to those observed in the active regions, while being restricted to semi-analytic axisymmetric solutions of the linear and nonlinear force-free field (NLFFF) equation were presented  in \citet{2013ASInC..10...51P} and \citet{2014ApJ...786...81P}.  In \citet{2014ApJ...786...81P} (hereafter PMR14), these solutions were used to simulate a library of photospheric vector magnetograms templates (depending upon the choice of parameters) that were compared with vector magnetograms observed by the spectro-polarimeter on board HINODE. {This technique is complimentary to the usual approach where the magnetograms are used as a boundary condition for a numerical NLFFF extrapolation \citep{2012LRSP....9....5W}. The solutions are first obtained on a local spherical shell and a planar surface is placed tangential to the inner sphere that represents a Cartesian cutout of an active region (see Figure 4 of PMR14 for more details). The orientation of the tangential plane are varied by two Euler rotations which are free parameters. The magnetic field calculated on this planar surface is then correlated with photospheric vector-magnetograms to fix the free parameters of the solutions. The radial component of magnetic field on the innermost shell is used to calculate the potential field for the volume of the shell. The three dimensional (3D) geometry of the magnetic field is used to estimate the relative helicity \citep{1984JFM...147..133B} and the free energy (difference in magnetic helicity and energy between the NLFFF and the corresponding potential field) for the entire volume of the shell. These values are then scaled with the solid angle subtended by the magnetogram to estimate the energetics of eruptive events like solar flares. The usefulness of this method is in obtaining fast and reasonably good fits to observed vector magnetograms using semi-analytical 3D NLFFF magnetic fields.

The rest of the paper is organized as follows. In \S \ref{s:cross}, we first present a description of the NLFFF solutions. The characterization of the amount of magnetic braiding for a toy model of two helices and for the various NLFFF solutions are presented in subsections \S \ref{helix} and \S \ref{nlff} using topological quantities like crossing and winding  numbers and their number distributions for different modes of the NLFFF solutions are also calculated. In \S \ref{bounds}, we discuss linking numbers and present estimates of the free energy and relative helicity for the field configurations, and also set bounds on their magnitude. Finally, the summary and conclusions are presented in \S \ref{conclude}.

\section{Calculation of crossing and winding for NLFFF solutions}
\label{s:cross}
The expression for the nonlinear force-free magnetic field in spherical geometry  follows from equation (36) of PMR14 is given by
\begin{equation}
 \mathbf{B}= \frac{-1}{r \sqrt{1-\mu^2}}\left(\frac{\sqrt{1-\mu^2}}{r}\frac{\partial \psi}{\partial\mu}\hat{\mathbf{r}}+
\frac{\partial \psi}{\partial r}\hat{\boldsymbol{\theta}}-Q\hat{\boldsymbol{\phi}}\right)
\label{low}
\end{equation}
where $\psi=(1- \mu^2)^{1/2} F(\mu)/r^n$, $Q=a\psi^{(n+1)/n}$, $a$ and $n$ are constants and $\mu=\cos\theta$. The above equation can also be obtained from equation (3) of \citet{1990ApJ...352..343L} by substituting for $\mu$. We can then write
\begin{equation}
\mathbf{B}= \left(\frac{-1}{r^{n+2}}\left[(1-\mu^2)^{1/2}F^\prime(\mu)-\frac{\mu F(\mu)}{(1-\mu^2)^{1/2}}\right],
\frac{n}{r^{n+2}}F,\frac{a}{r^{n+2}}(1-\mu^2)^{1/2n}F^{1+1/n}\right)
\label{flow}
 \end{equation}
where $F$ is obtained from 
\begin{equation}
 (1-\mu^2)F^{\prime\prime}(\mu)-2\mu F^\prime(\mu)+\left[n(n+1)-\frac{1}{(1-\mu^2)}\right]
F(\mu)+a^2\frac{(n+1)}{n}F^{\frac{(n+2)}{n}}(1-\mu^2)^\frac{1}{n}=0,
\label{feq}
\end{equation}
which has to be solved numerically as an eigenvalue problem for the variable $a$ for a given value of $n$. For $n=p/q$, where $p$ and $q$ are integers prime to each other and $q\neq 0$, solutions exist for all odd values of $p$, while for even values of $p$, it exists only if $F(\mu)>0$ in the domain $-1 \leq \mu \leq 1$ (PMR14). 
 \begin{figure}[h!]
\centering
\includegraphics[scale=.72]{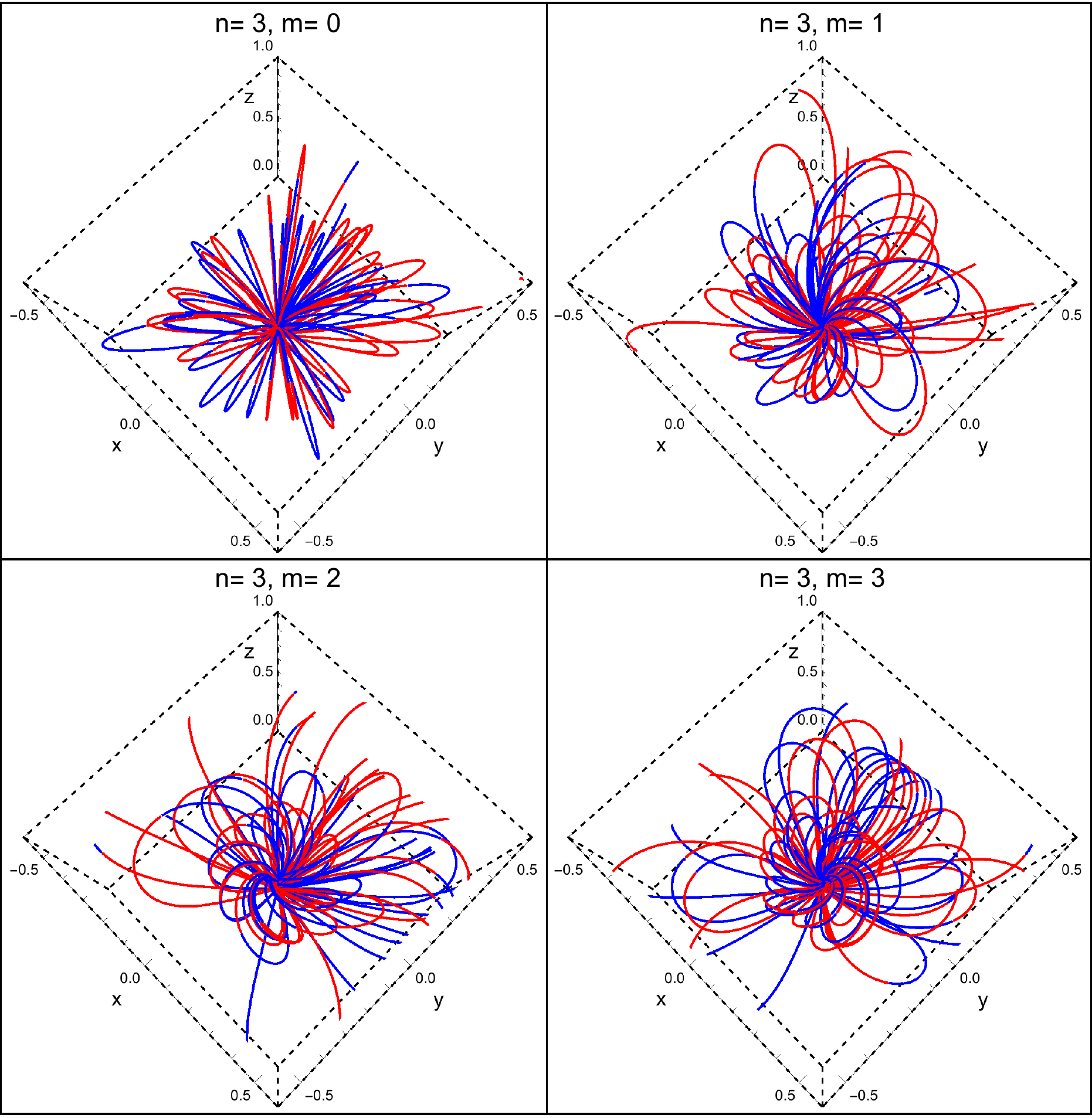}
\caption{Plots depicting magnetic field lines for the axisymmetric NLFFF modes corresponding to $n=3$ and $m=0-3$ with 50 randomly sampled seed points in the domain $-0.5<x<0.5$, $-0.5<y<0.5$, $z=0.5$ . The blue (red) color corresponds to the field line segment where $B_z$ is positive (negative).  }
\label{nmpltf}
\end{figure}
The magnetic field lines of the solutions for $n=3$ and $m=0-3$ (which correspond to different eigenvalues of $a$ in equation \eqref{feq}) are shown in Figure \ref{nmpltf}. The plots are shown in a Cartesian domain (following the convention of \citet{1990ApJ...352..343L}, where the point sources are assumed to be located at the origin and the fields are calculated for the region $z>0$) for 50 field lines that are sampled randomly between $-0.5<x<0.5$, $-0.5<y<0.5$ and $z=0.5$. No preference has been given to the strength of the magnetic field while choosing the location of the points. The field lines are then obtained from a bi-directional integration of the equation of field lines between the planes $z=0.005$ and $z=1$. The mode $m=0$ corresponds to potential fields, which represent the simplest untwisted geometry for the field lines, whereas $m>0$ modes represent twisted fields. The blue and red color field line segments in Figure \ref{nmpltf} correspond to positive and negative values of $B_z$ respectively. It is clearly seen in the figure that in the case of $m=0$, for each field line, there is a plane that completely confines it, whereas, for higher values of $m$, the field lines are no longer confined to a single plane which allows for the possibility of two oppositely directed field lines coming in close proximity, which in turn is a favorable scenario for small-scale reconnections (Parker's theory of nanoflare heating \citep{1983ApJ...264..642P}).

In order to study the braiding between the field lines, \citet{1993PhRvL..70..705B} proposed to use the concept of crossing numbers (previously discussed by \citet{calugareanu1959integrale,freedman1991divergence,1992RSPSA.439..411M}) that can be understood as follows. Consider two field lines stretching between two planes $z=0$ and $z=L$. Let $\phi'$ be the polar angle in the x-y plane (see Figure \ref{cross2}). When the curves are observed from a viewing angle $\phi'$ the two curves will exhibit a certain number of crossovers, $c(\phi')$. 
To compute the crossing number for two curves $\mathbf{x_1}(z)$ and $\mathbf{x_2}(z)$, where $\mathbf{x_1}(z)=(x_1,y_1)$, we write the displacement vector as $\mathbf{r_{12}}(z) = \mathbf{x_2}(z)-\mathbf{x_1}(z)$, which makes an angle $\theta_{12}(z)$ with the $z$ axis, such that  $\displaystyle{\theta_{12}= \arccos\left(\frac{\mathbf{r_{12}}}{r_{12}}\cdot \hat{\mbf{x}}\right)}$, where $\hat{\mbf{x}}$ is the position vector along the $x$ axis. The crossing number for these curves can then be written as \citep{1993PhRvL..70..705B}
\begin{equation}
c=\frac{1}{\pi}\int_0^L\left|\frac{\dif \theta_{12}}{\dif z}\right|\dif z.
\label{ecross}
\end{equation}
\begin{figure}[h!]
\centering
\includegraphics[scale=.5]{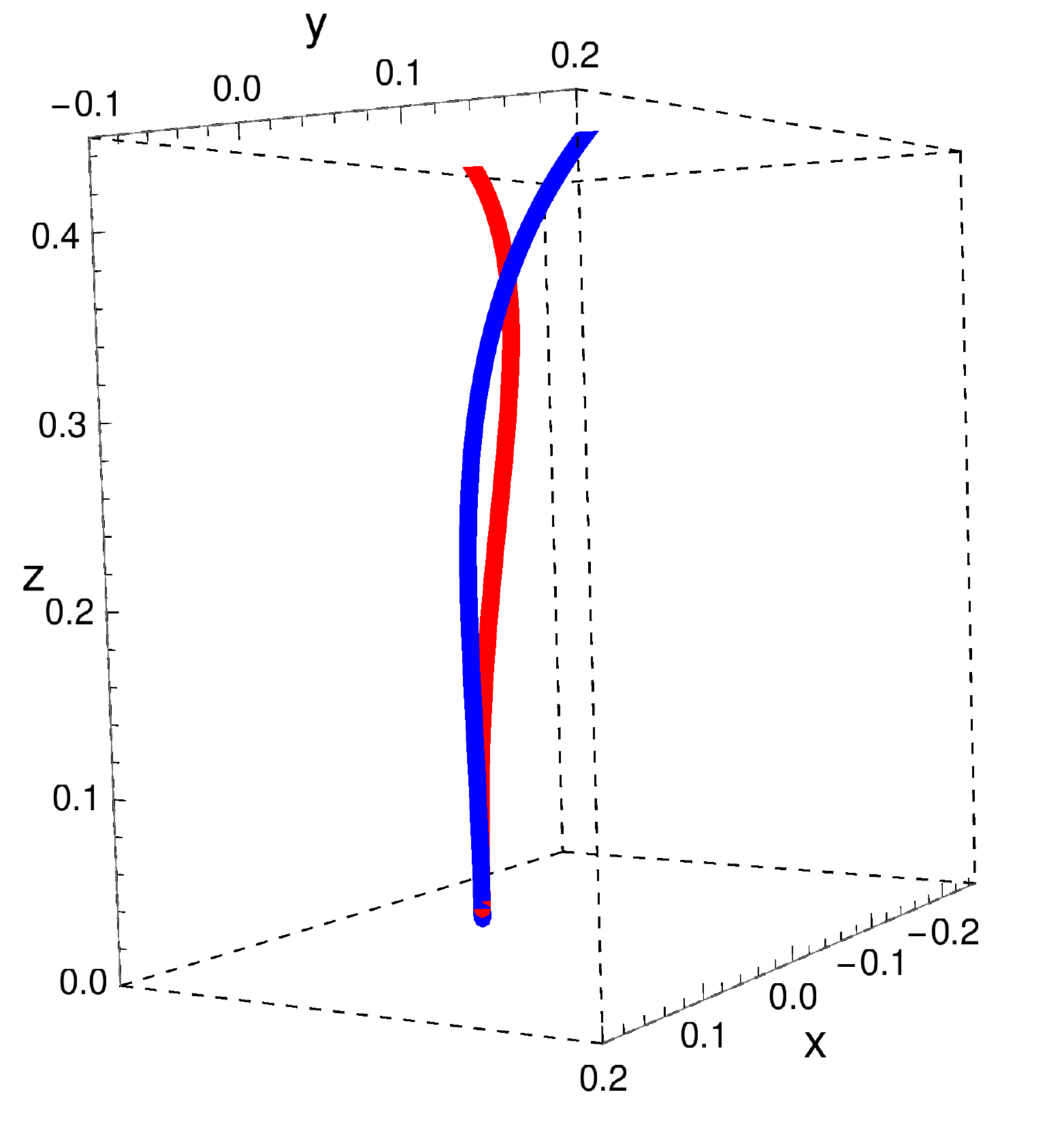}
\caption{Two field lines corresponding to modes ($n=3,~m=2$) between two planes $z=0$ and $z=0.45$ exhibit  a crossover. The upper boundary is chosen such that there are no reversals of $B_z$ in either case, so that a crossing number can be calculated.}
\label{cross2}
\end{figure}
The crossing number is not a topological invariant, but a proxy for magnetic energy; for a given field line topology, it has a positive minimum value that is viewing angle dependent.
An alternative way to estimate this braiding is to calculate the winding numbers for the magnetic field lines which is defined as follows. Let $\mbf{B_1}$ and $\mbf{B_2}$ define two magnetic field lines where the corresponding line elements are given by $\mathbf{r_1}$ and $\mbf{r_2}$. We can write
\begin{equation}
\mathbf{r_1}(z+ \dif z)=\mbf{r_1}(z)+\mbf{n_1} \dif s,
\end{equation}
where $\disp{\mbf{n_1}=\frac{\mbf{B_1}}{B_1}=\frac{\mbf{B_{1z}}+\mbf{B_{1t}}}{B_1}}= (\hat{\mbf{z}} + \mbf{t_1})/ \sqrt{1+t_1^2}$ is the unit vector along $\mbf{B_1}$, and $\mbf{t_1}=\mbf{n_1}-(\mbf{n_1}\cdot \hat{\mbf{z}})$ is a vector along the transverse component of $\mbf{B_1}$; $\mbf{B_{1z}}$ and $\mbf{B_{1t}}$ are the longitudinal and transverse components of $\mbf{B_1}$. Here $t_1= B_{1t}/B_{1z}$,  $\dif s$ is the infinitesimal displacement along the field line, and $\disp{\frac{\dif s}{\dif z}=\frac{B_1}{B_{1z}}}$ gives the equation of the field line, so that
\begin{equation}
\displaystyle \dif \mbf{r_1}=\mbf{n_1}(z)\frac{B_1}{B_{1z}}\dif z =\frac{(\hat{\mbf{z}} + \mbf{t_1})}{ \sqrt{1+t_1^2}} \frac{B_1}{B_{1z}}\dif z=(\mbf{t_1}+\hat{\mbf{z}})\dif z.
\end{equation}
We define the twist energy as the magnetic energy associated with the transverse field as $E_t = V B_t^2/(8 \pi)$ and
the twist vector, $\mbf{t_2} -\mbf{t_1}$ as the difference between the transverse components of two
field lines under consideration. The twist angle $\phi_{12}$ between the field lines can now be expressed as \citep{1971PNAS...68..815B}
\begin{equation}
\dif \phi_{12} \hat{\mbf{z}}=\frac{\dif \mbf{r_{12}}}{r_{12}}\times \hat{\mbf{r}}_{12}=\frac{1}{|\mbf{r_1}-\mbf{r_2}|}\dif\left(\mbf{r_2}(z)-\mbf{r_1}(z)\right)\times \hat{\mbf{r}}_{12}.
\end{equation}
Thus, we find 
\begin{align}
\frac{\dif \phi_{12}}{\dif z} \hat{\mbf{z}}&=\frac{1}{r_{12}^2}\left(\frac{\dif \mbf{r_2}}{\dif z}-\frac{\dif \mbf{r_1}}{\dif z}\right)\times\mbf{r_{12}}\nonumber\\
&=\left(\mbf{n_2}\frac{B_2}{B_{2z}}-\mbf{n_1}\frac{B_1}{B_{1z}}\right)\times\frac{\hat{\mbf{r}}_{12}}{r_{12}}.
\label{phi12}
\end{align}

Next, we show the difference between the crossing number $c$ \citep{1993PhRvL..70..705B} and the winding number $w$ that we use here. First we write
\begin{align}
\frac{\dif \phi_{12}}{\dif z} &= \left (\mbf{n_2} \sqrt{1+t_2^2} -\mbf{n_1} \sqrt{1+t_1^2} \right) \times \frac{\hat{\mbf{r}}_{12}}{r_{12}} \cdot \hat{\mbf{z}} \nonumber \\ 
&= {1 \over r_{12}} \left( \mbf{t_2} -\mbf{t_1} \right ) \cdot \hat{\boldsymbol{\phi}}_{12},
\end{align}
where $ \hat{\boldsymbol{\phi}}_{12}=\hat{\mbf{z}} \times \hat{\mbf{r}}_{12}$.  Finally, the winding number can be defined as 
\begin{equation}
w=\frac{1}{\pi}\int_0^L \left|\frac{\dif \phi_{12}}{\dif z}\right| \dif z.
\label{e:crs}
\end{equation}

 In \citet{1993PhRvL..70..705B}, the vector $\mbf{t_2} -\mbf{t_1}$ was further simplified by taking $B_{1z}= B_{2z}=B_0$. 
We elaborate on the differences between this and our approach:
\begin{enumerate}
\item Firstly,  in the case  when equation \eqref{ecross} is used, $\theta_{12}$ represents the winding of the line segment $\mbf{r}_{12}(z)$ at a given height  with respect to a fixed $x$ direction, while in our case,  equation \eqref{e:crs} is used and $\phi_{12}$
represents the winding of the two field lines about each other regardless of the direction chosen for $x$. Now, the
two would be equal only if $\theta_{12}$ is viewing angle averaged. 

\item Secondly, our application of the formula for the winding number is for a general case, where both the
vertical and the tangential components vary, while in the calculation by \citet{1993PhRvL..70..705B}, the vertical field is taken to be uniform while the tangential field fluctuates.
\end{enumerate}

\subsection{Calculations of crossing numbers for the case of helices}
\label{helix}
\begin{figure}[h!]
  \centering
  \begin{subfigure}[]{0.45\textwidth}
    \centering
    \includegraphics[width=0.7\linewidth]{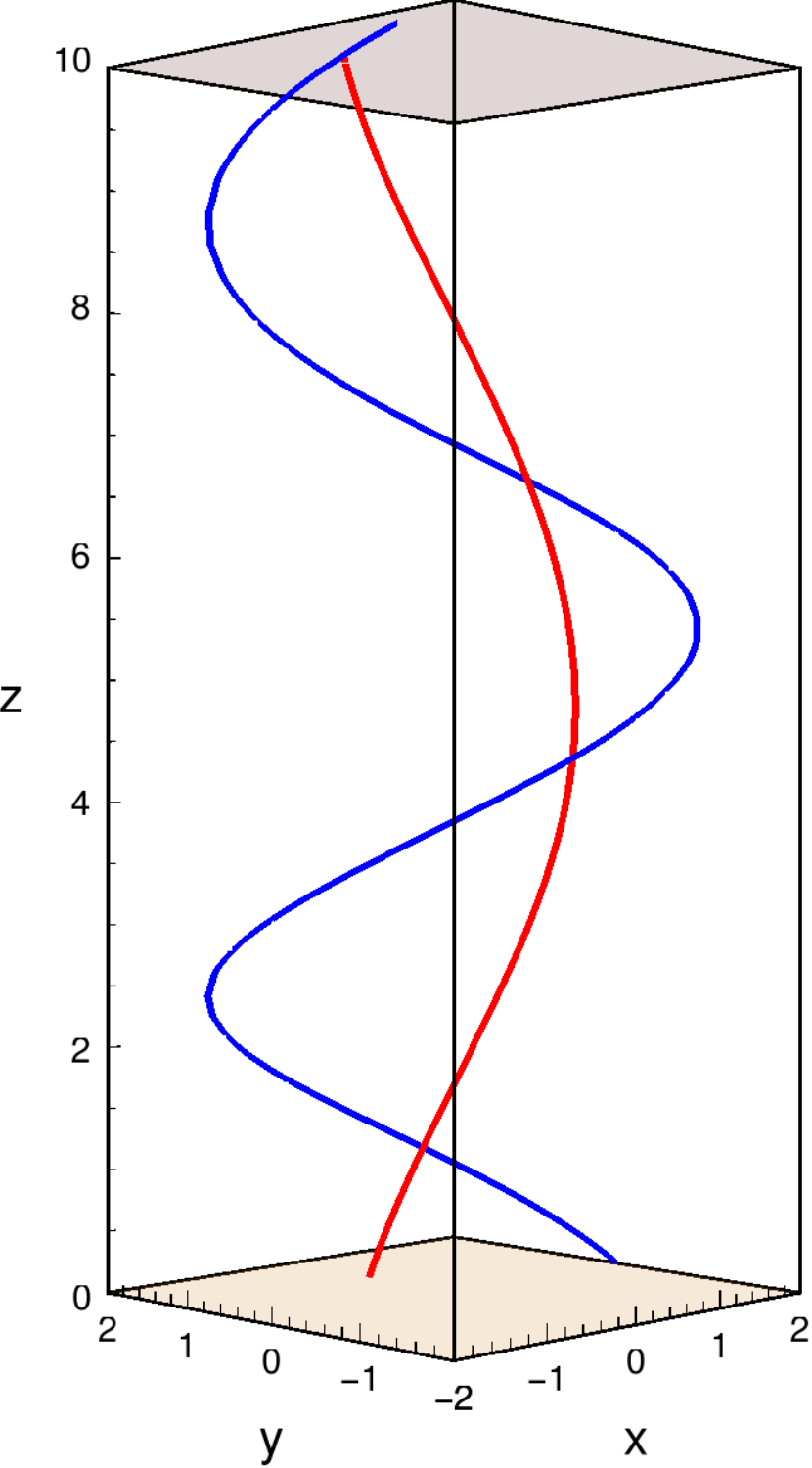}
    \caption{}
  \end{subfigure}
\quad
  \begin{subfigure}[]{0.45\textwidth}
    \centering
    \includegraphics[width=0.75\linewidth]{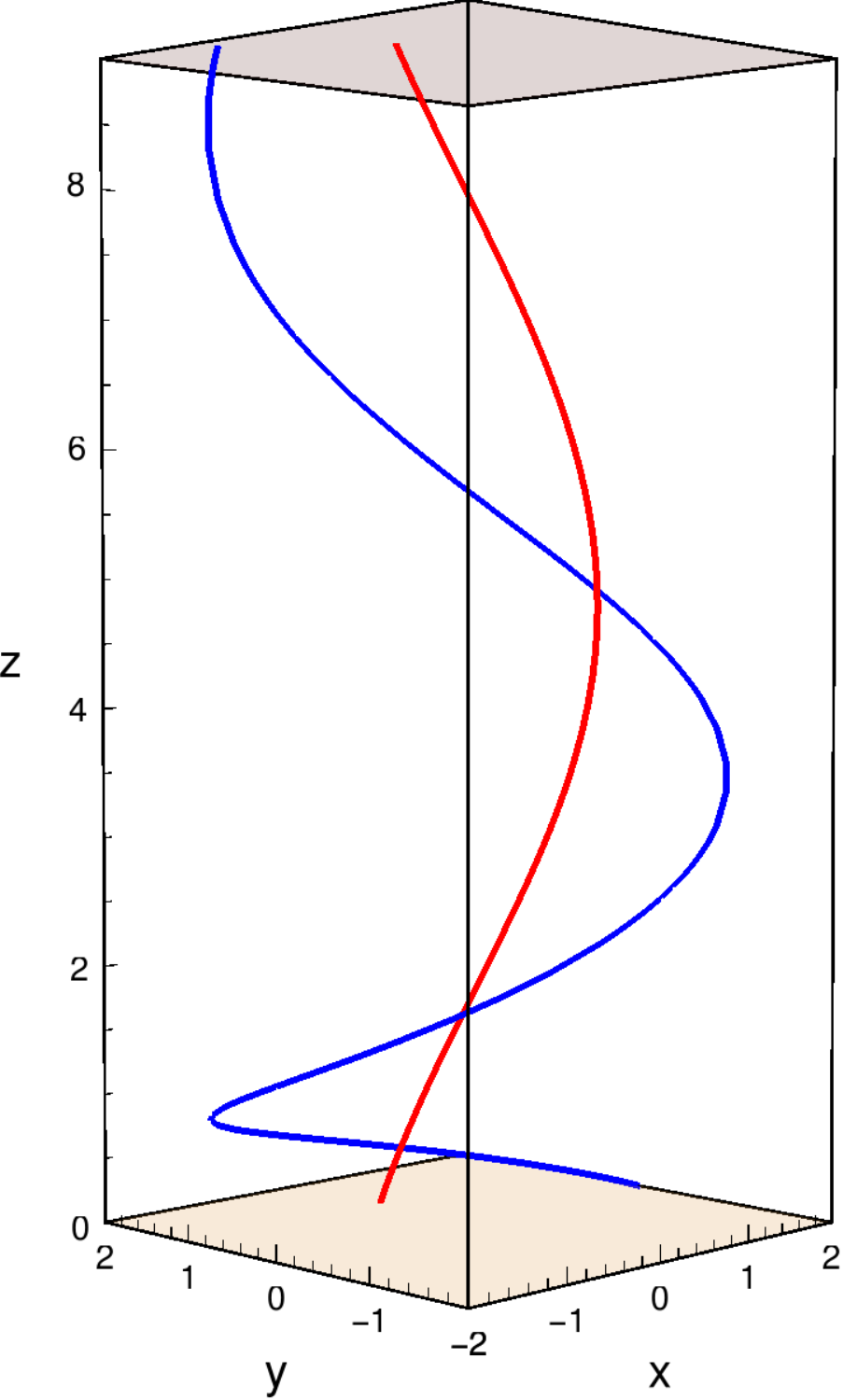}
    \caption{}
  \end{subfigure}
   \centering
  \begin{subfigure}[]{0.45\textwidth}
    \centering
    \includegraphics[width=1\linewidth]{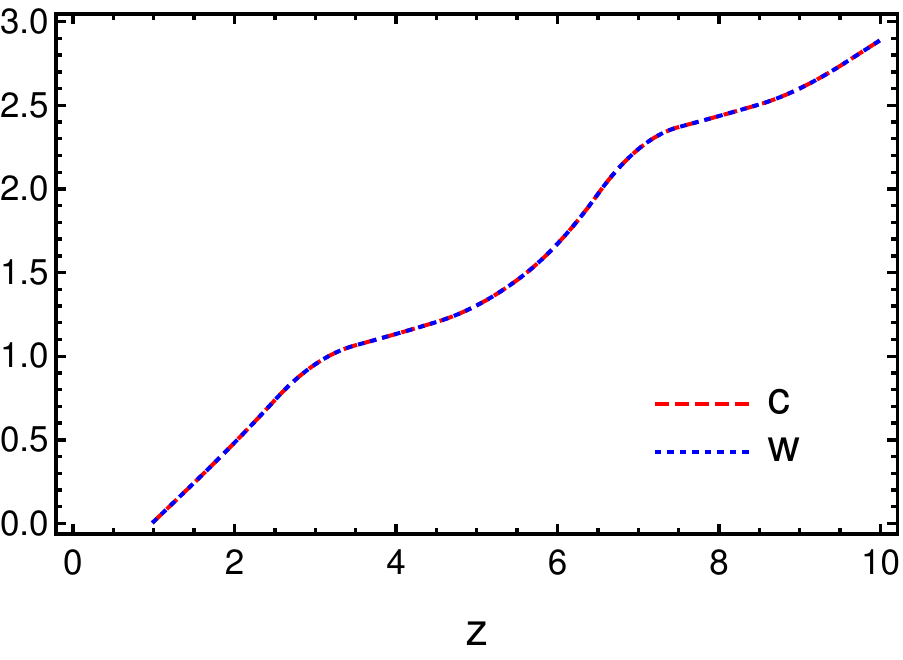}
    \caption{}
  \end{subfigure}
\quad
  \begin{subfigure}[]{0.45\textwidth}
    \centering
    \includegraphics[width=1\linewidth]{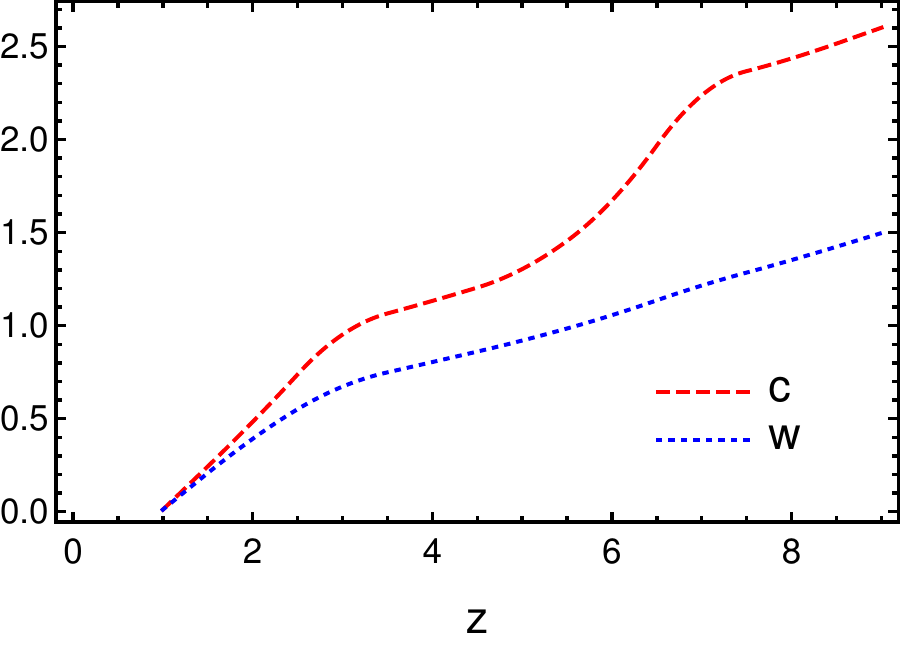}
    \caption{}
  \end{subfigure}
  
    \caption{Panels (a) and (b) represent the parametric plots for pairs of curves represented by equations \eqref{e:case1} and \eqref{e:case2} respectively. The corresponding variation in $c$ and $w$ with $z$ for these cases are shown in panels (c) and (d).}
  \label{cross}
\end{figure}
 
In order to explore the differences between the two formulae, we explore two cases:
\begin{enumerate}[label=(\Alph*)]
\item We consider two curves $\mbf{r_1}$ and $\mbf{r_2}$ (defined  in Cartesian) with the following parametric dependence on $t$: 
\begin{equation}
\mbf{r_1}=\left(\sin(t/2),\cos(t/2),t\right),  \quad \mbf{r_2}=\left(2\cos(t),2\sin(t),t\right).
\label{e:case1}
\end{equation}
We see that the $z$ variation for both the curves are linear which would imply magnetic field lines with constant $B_z$. These curves are shown in Figure \ref{cross}(a). The variation of the quantities $c$ and $w$ is shown in Figure \ref{cross}(c). We find that both these quantities behave in the same manner with increasing $z$.
\item Now we introduce a quadratic growth of $z(t)$ for one of the curves (which corresponds to varying $B_z$ case), such that
\begin{equation}
\mbf{r_1}=\left(\sin(t/2),\cos(t/2),k_1 t\right),  \quad \mbf{r_2}=\left(2\cos(t),2\sin(t),k_2 t + k_3 t^2\right).
\label{e:case2}
\end{equation}
The parametric plot for these curves with $k_1=1$, $k_2=0.1$ and $k_3=0.1$ is shown in Figure \ref{cross}(b). From Figure \ref{cross}(d), we find that $c$ and $w$ now differ appreciably. Thus we can conclude that both $c$ and $w$ should considered while studying braiding for field lines with varying $B_z$.
\end{enumerate}
In the non-linear force-free topology that we consider, we find that there is a considerable variation of $B_z$ over a small region. Hence, there is a need to study the distributions for both $c$ and $w$; this is presented in the next subsection.

\subsection{Calculations of crossing numbers for the NLFFF solutions}
\label{nlff}
The distributions of crossing and winding numbers based on equations \eqref{ecross} and \eqref{e:crs} for the NLFFF modes are shown in Figure \ref{hist}. In order to calculate these numbers, we consider all possible pairs from the field lines shown in Figure \ref{nmpltf}. For each pair, we first calculate the height at which the vertical field reverses for both the field lines. Based on this, we choose a common minimum height, such that both the field lines can be uniquely defined between these two planes, and the topological quantities are then calculated in this domain.
\begin{figure}[h!]
  \centering
  \begin{subfigure}[]{0.45\textwidth}
    \centering
    \includegraphics[width=1\linewidth]{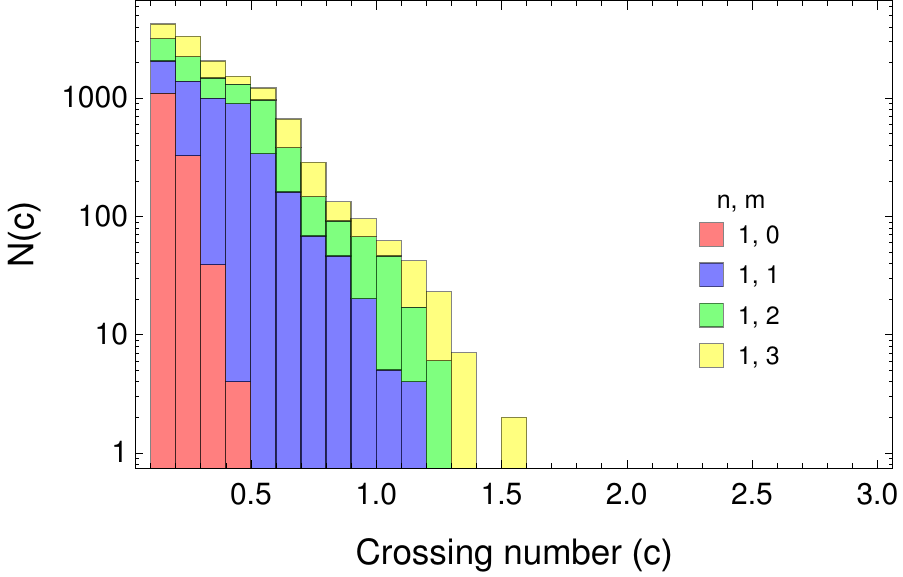}
    \caption{}
  \end{subfigure}
\quad
  \begin{subfigure}[]{0.45\textwidth}
    \centering
    \includegraphics[width=1\linewidth]{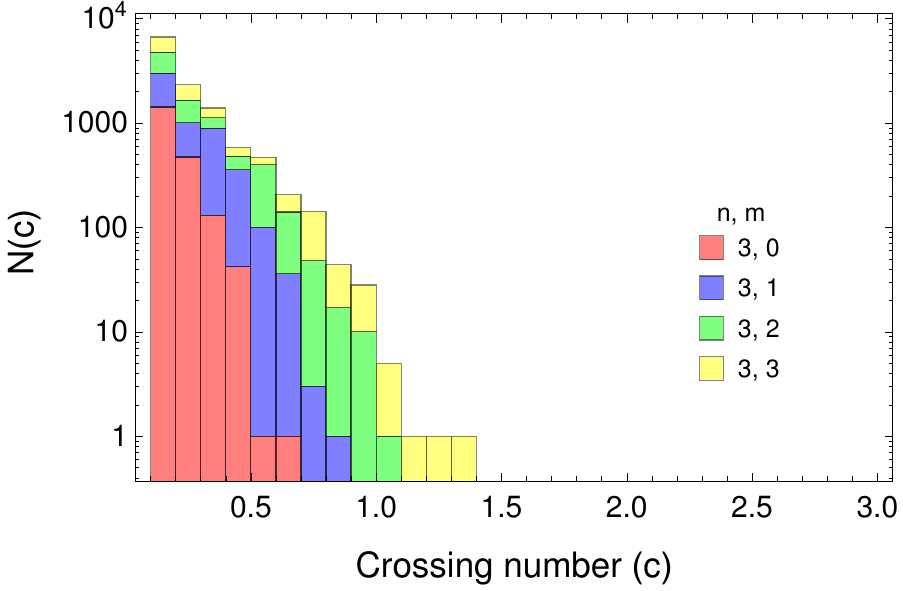}
    \caption{}
  \end{subfigure}
   \centering
  \begin{subfigure}[]{0.45\textwidth}
    \centering
    \includegraphics[width=1\linewidth]{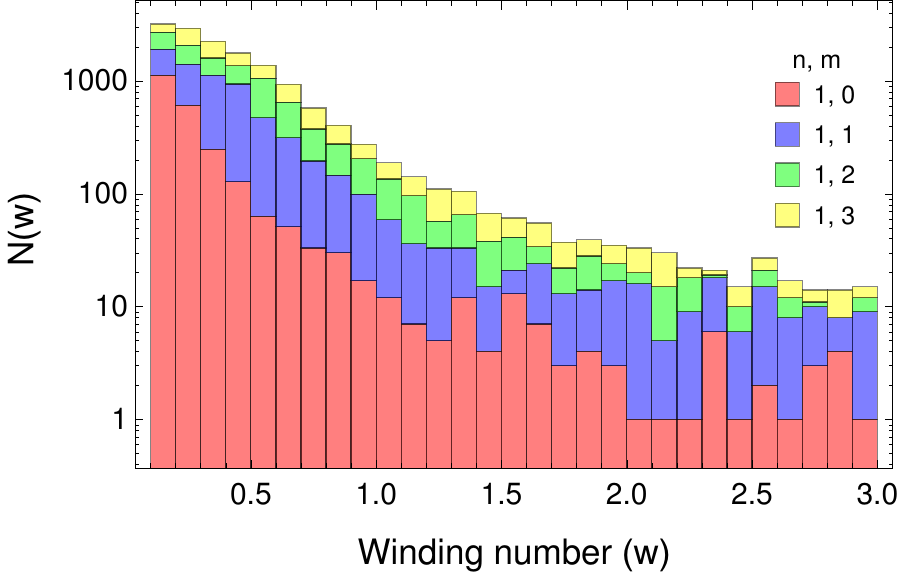}
    \caption{}
  \end{subfigure}
\quad
  \begin{subfigure}[]{0.45\textwidth}
    \centering
    \includegraphics[width=1\linewidth]{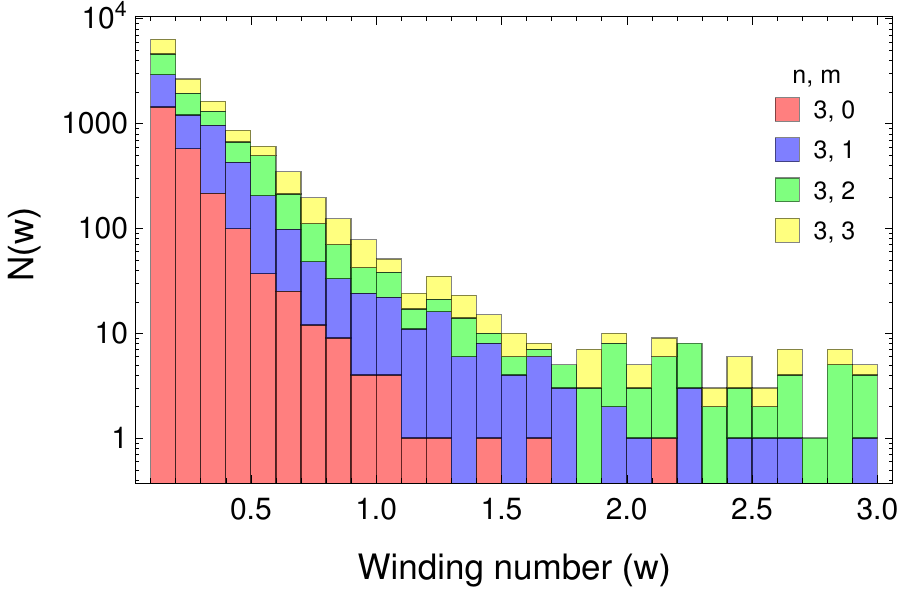}
    \caption{}
  \end{subfigure}
  \begin{subfigure}[]{0.45\textwidth}
    \centering
    \includegraphics[width=1\linewidth]{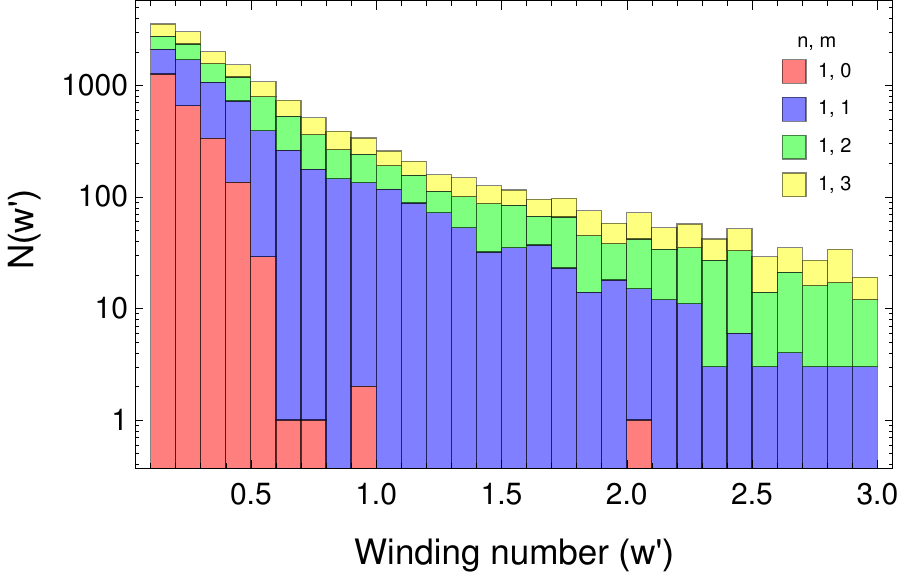}
    \caption{}
  \end{subfigure}
\quad
  \begin{subfigure}[]{0.45\textwidth}
    \centering
    \includegraphics[width=1\linewidth]{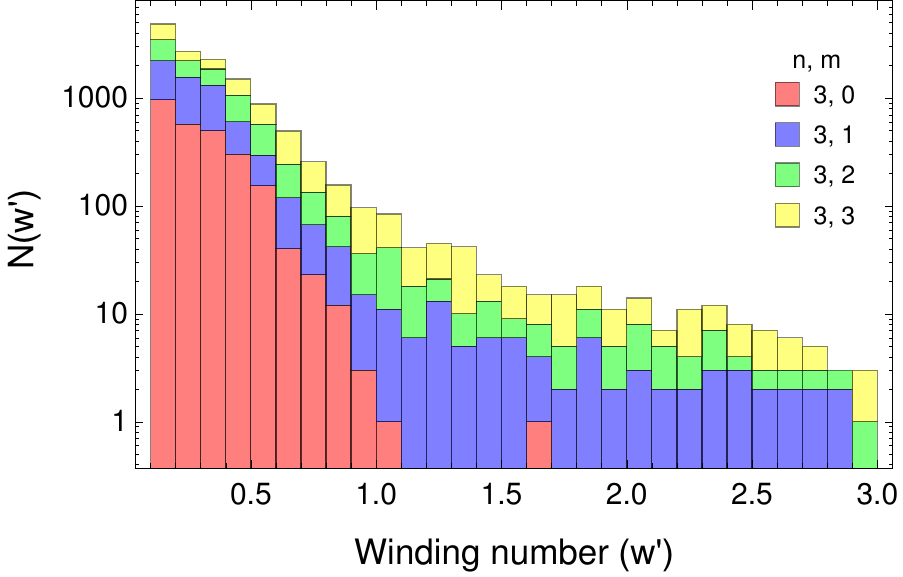}
    \caption{}
  \end{subfigure}
    \caption{Stacked histograms depicting the distribution of number of pairs $N$ for a given values of crossing number $c$ (panels (a) and (b)) and winding number $w$ (panels (c) and (d)) for $n =1$, $m = 0-3$ and $n=3$, $m=0-3$ modes. Panels (e) and (f) depict the same for winding  $w'$, which is calculated with the approximation $B_{1z}=B_{2z}=B_0$. The field lines considered for pairing are same as those shown in Figure \ref{nmpltf}.}
  \label{hist}
\end{figure}
In order to explore the differences between the two formulae, we set up two experiments: one where we calculate the winding number using equation \eqref{e:crs} and compare it with the crossing number using equation \eqref{ecross} to see the 
impact of a fixed viewing angle, while keeping the field topology the same and another to emphasize the difference between the two, by calculating the winding number (by using the varying $B_z$ in equation \eqref{phi12}) and comparing it with the winding number (by using an average $B_0= B_{z1}=B_{z2}$ in equation \eqref{phi12}).
In Figure \ref{hist}, the stacked histograms depicting the distribution of number of pairs $N$ (shown in log scale) are presented for the crossing $c$ (panels (a) and (b)) and winding number $w$ (panels (c) and (d)) for $n =1$, $m = 0-3$ and $n=3$, $m=0-3$ modes. We find that values for $c$ are predominant in the range of $0-1.5$, whereas $w$ has significant distribution throughout the range $0-3$. This difference can be attributed to the effect of viewing-angle dependence of $c$. To estimate the effect of the approximation of $B_{1z}=B_{2z}=B_0$, we plot the distribution of winding numbers calculated with this approximation ($w'$) in panels (e) and (f) of Figure \ref{hist}. The clear difference seen in the distributions of $w$ and $w'$ highlights the importance of $B_z$ dependence in the calculations of crossing numbers. To bring out the difference between $c$ and $w$ more clearly, we show the scatter plots between $c$ and $w$ in Figure \ref{f:sp2}, where the Pearson $r$ coefficient is also mentioned. We find that the correlation is better for lower values of $m$. We calculate the cumulative distribution functions (CDFs) and perform the Kolmogorov-Smirnov (KS) test on the two data sets for $c$ and $w$ (Table \ref{t:ksl}). Plots of CDFs for two illustrative cases corresponding to $n,~m=1,0$ and $n,~m=3,2$ are shown in Figure \ref{f:comp}. Based on the KS test statistics given in Table \ref{t:ksl}, we reject the null hypothesis that the data sets have the same distribution at the 5 \% level.  Further, the $p$ value in all the cases was found to be zero to the accuracy of $10^{-10}$. This means that the two distributions of the crossing and winding numbers can be treated as statistically different.

\begin{figure}[h!]
\centering
\includegraphics[scale=.75]{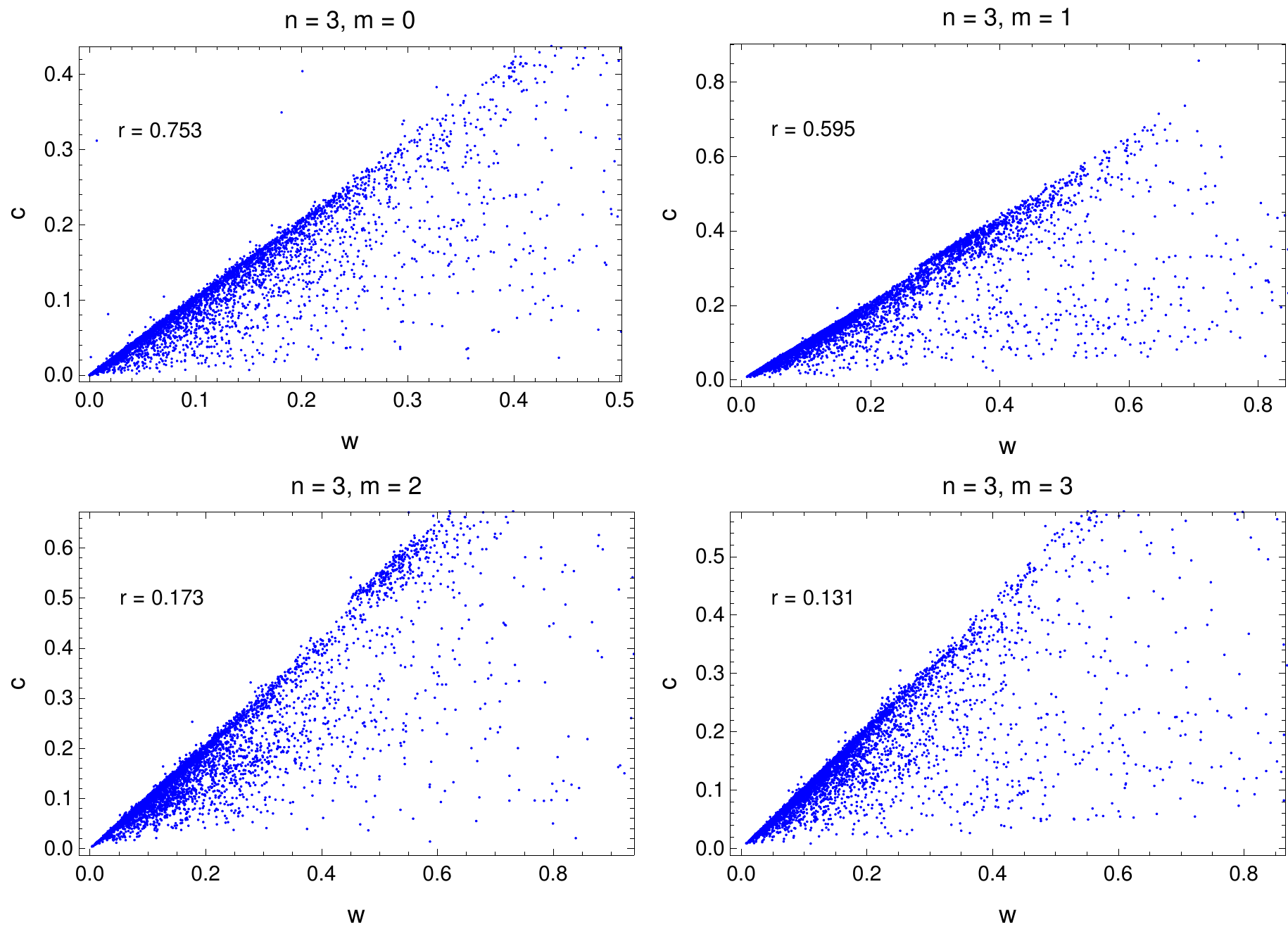}
\caption{Scatter plots comparing distributions of $c$ and $w$ for $n=3$ and $m=0-3$. The Pearson $r$ coefficient is also mentioned for each case. }
\label{f:sp2}
\end{figure}

\begin{table}[h!]
\centering
\resizebox{0.7\textwidth}{!}{%
\begin{tabular}{|c|c|c|c|c|c|c|c|c|}
\hline
n  & 1     & 1     & 1     & 1     & 3     & 3     & 3     & 3     \\ \hline
m  & 0     & 1     & 2     & 3     & 0     & 1     & 2     & 3     \\ \hline
KS & 0.291 & 0.150 & 0.156 & 0.217 & 0.105 & 0.074 & 0.090 & 0.094 \\ \hline
\end{tabular}%
}
\caption{Test statistic values for the Kolmogorov-Smirnov (KS) test performed on the crossing and winding number distributions for different values of $n$ and $m$. The $p$ value in all the cases was found to be zero to the accuracy of $10^{-10}$.}
\label{t:ksl}
\end{table}

\begin{figure}[h]
  \centering
  \begin{subfigure}[]{0.45\textwidth}
    \centering
    \includegraphics[width=1\linewidth]{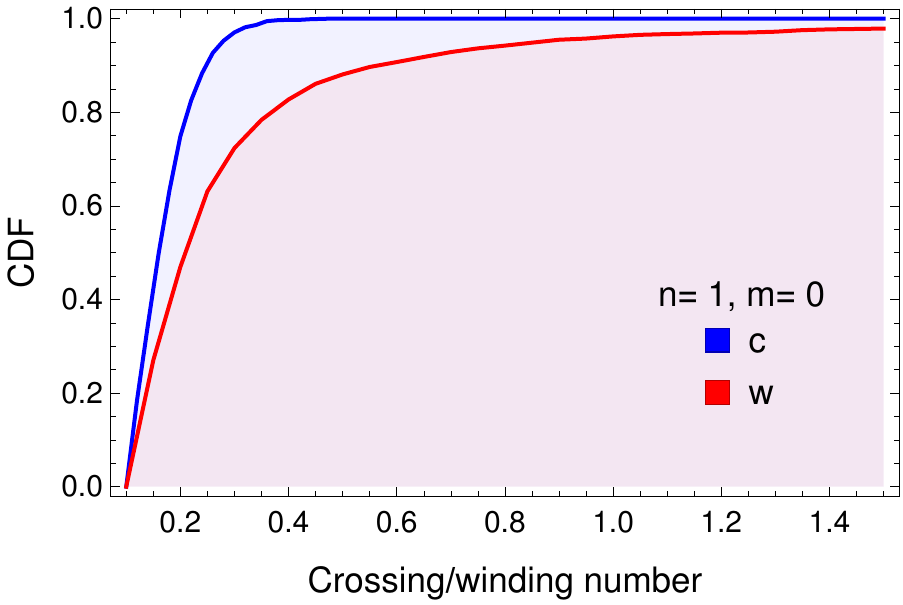}
    \caption{}
    \label{n1m}
  \end{subfigure}
\quad
  \begin{subfigure}[]{0.45\textwidth}
    \centering
    \includegraphics[width=1\linewidth]{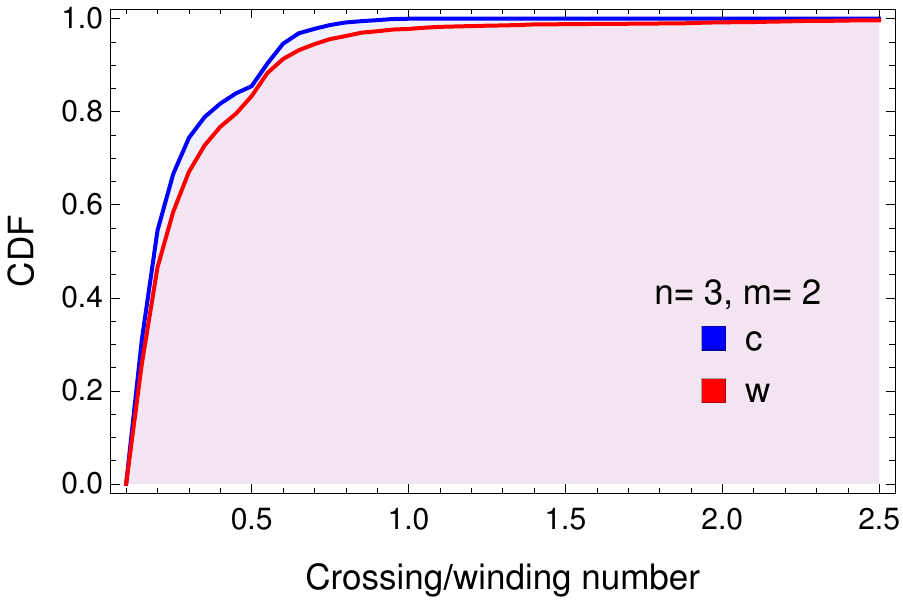}
    \caption{}
    \label{n3m}
  \end{subfigure}
 \caption{Comparison of the cumulative distribution functions (CDFs) of the crossing and winding number distributions for (a) $n =1$, $m = 0$ (b) $n=3$, $m=2$ modes to illustrate that the two descriptions of numbers $c$ and $w$ are statistically different.}
\label{f:comp}
\end{figure}

\section{Linking numbers and energy bounds for NLFFF solutions from braiding}
\label{bounds}
In this section, we make an estimate of the free energy using the definition of equation \eqref{e:crs} which is more appropriate as it represents the twist energy. We estimate the global winding number summed over the volume due to braiding to be \citep[cf.][]{1993PhRvL..70..705B}
\begin{equation}
W= {1 \over 2 \pi} \int_0^L \int \int B_{1z} B_{2z} \left|\frac{\dif \phi_{12}}{\dif z}\right| \dif^2 x_1 \dif^2 x_2 \dif z,
\label{e:W}
\end{equation} 
and it follows that 
\begin{align}
{\dif W \over \dif z}&=  \int \int {B_{1z} B_{2z} \over  2 \pi r_{12}} \left|\left( \mbf{t_2} -\mbf{t_1} \right ) \cdot \hat{\boldsymbol{\phi}}_{12} \right| \dif^2 x_1 \dif^2 x_2 \\ 
&\leq \int \int {B_{1z} B_{2z} \over  \pi r_{12}} \left|\left( \mbf{t_1}  \cdot \hat{\boldsymbol{\phi}}_{12} \right ) \right| \dif^2 x_1 \dif^2 x_2 \nonumber \\
&= {1 \over \pi} \int  B_{1t} \sigma_1 \dif^2 x_1 \int  { \left|\left( \hat{\mbf{t}}_1 \cdot \hat{\boldsymbol{\phi}}_{12}\right ) \right| B_{2z} \over   r_{12}} \dif^2 x_2  \nonumber \\
&\leq {B_0 \over \pi} \int  B_{1t} \sigma_1 \dif^2 x_1 \left [ \int  { \left|\left( \hat{\mbf{t}}_1  \cdot \hat{\boldsymbol{\phi}}_{12}\right ) \right| \over   r_{12}} \dif^2 x_2  + \int {|b_{2z}|\over   r_{12}} \dif^2 x_2 \right ] \nonumber \\
& \equiv {B_0 \over \pi} \int  B_{1t} \sigma_1 \dif^2 x_1 \left [ {\cal I}_1 +  \int {|b_{2z}|\over   r_{12}} \dif^2 x_2 \right ], 
\label{e:dwdz}
\end{align}
where $\sigma_1$ represents the sign of $B_{1z}$ and $B_{2z}= B_0+ B_0 b_{2z}$ so that  $b_{2z}= (B_{2z}-B_0)/B_0$ represents the normalized fluctuating part of $B_{2z}$, where $B_0$ is a constant (mean vertical field) and $B_0 b_{2z}$ is the varying vertical field. Likewise the quantities $b_{1z} \equiv (B_{1z}-B_0)/B_0, b_{1t} \equiv B_{1t}/B_0, b_{2t} \equiv B_{2t}/B_0$ are thus defined. The second line in equation \eqref{e:dwdz} uses the triangle inequality

\begin{equation}
|(\mbf{t_2}-\mbf{t_1})\cdot \hat{\phi}_{12}| \leq |\mbf{t_2}\cdot \hat{\phi}_{12}|+|\mbf{t_1}\cdot \hat{\phi}_{12}| \equiv t_{1\phi} + t_{2\phi} \leq 2 \mathrm{Max[}t_{1\phi}, t_{2\phi}]\equiv 2 t_{1\phi}. 
\end{equation}

Also, the quantity ${\cal I}_1$ is given by
\begin{equation}
\mathcal{I}_1 = \int  { \left|\left( \hat{\mbf{t}}_1  \cdot \hat{\boldsymbol{\phi}}_{12}\right ) \right| \over   r_{12}} \dif^2 x_2 \equiv R f(a_1),  
\end{equation}

where $a_1= x_1/R$ while taking a cylindrical cross section of radius $R$ for the flux. Applying the Cauchy-Schwarz inequality twice, we find after rearranging the terms that
\begin{align}
{\dif W \over \dif z}& \leq {B_0^2 \over \pi} \left (\int b_{1t}^2 \dif^2 x_1\right)^{1/2} 
\left[ \left (\int {\cal I}_1^2 \dif^2 x_1\right)^{1/2} + \left (\int b_{2z}^2 \dif^2 x_2\right)^{1/2}  \left( \int \int { 1 \over r_{12}^2} \dif^2 x_2 \dif^2 x_1 \right)^{1/2} \right], \nonumber \\
&\equiv B_0^2  R^3 \epsilon_t^{1/2} \left [{\cal J}_1^{1/2} + \epsilon_z^{1/2} {\cal I}_2^{1/2} \right],
\label{e:dwdz2}   
\end{align}
where the mean dimensionless twist defined by $\disp b_{1t}^2 \equiv \frac{B_{1t}^2}{B_0^2}= \frac{B_{1t}^2}{B_{1z}^2} \frac{B_{1z}^2}{B_0^2} = t_1^2(1+b_{1z}^2)$ and
the corresponding mean values in the cross section are
\begin{equation}
\epsilon_t =  \frac{1}{\pi R^2}\int b_{1t}^2 \dif^2 x_1 \equiv <b_{1t}^2 > ~~~~\mathrm{and} ~~~~~\epsilon_z = \frac{1}{\pi R^2} \int b_{2z}^2 \dif^2 x_2 \equiv <b_{2z}^2 >.
\end{equation}
The integrals ${\cal J}_1$ and ${\cal I}_2$ are defined to be
\begin{equation}
{\cal J}_1= \frac{1}{\pi R^4} \int {\cal I}_1^2 \dif^2 x_1= 2 \int_0^1 f^2 (a_1) a_1 \dif a_1  \equiv \ol{f^2}
\end{equation}
\begin{equation}
{\cal I}_2 = \int \int { 1 \over r_{12}^2} \dif^2 x_2 \dif^2 x_1 \equiv  R^2 \kappa
\end{equation} 
 With these definitions
and substituting the twist energy $E_t = V B_t^2/(8 \pi)= \pi B_0^2 R^2 L \epsilon_t / (8 \pi)$ in
equation \eqref{e:dwdz2} and integrating it to a length $L$ and squaring, we find that the lower bound for $E_t$ is given by
\begin{equation}
E_t \geq {W^2 \over 8 R^4 B_0^2 L g^2} 
\label{Etbound}
\end{equation}
where $g$ is given by
\begin{equation}
g= \ol{f^2}^{1/2} + \sqrt{\kappa} \epsilon_z^{1/2},
\end{equation}
and we estimate numerically that $g \approx 4$ for $\epsilon_z=0.01$ given $\kappa=13.68$ and  $\ol{f^2}=13.14$ (see Appendix for a sketch of the calculations of these estimates).  We plan to present further details of the analytic calculations for $\ol{f^2}$ and $\kappa$ integrals and verify the resulting bounds  for various topologies in a paper in preparation.

The calculation of crossing number further allows us to estimate a proxy of the relative helicity (defined as a sum of the linkages, \citet{1986GApFD..34..265B}) in this domain arising from braiding of the magnetic field lines in the following manner 
\begin{equation}
L_r=(2\pi)^{-1}\int \int_{z=0}B_z(\mathbf{x})B_z(\mathbf{x}') \delta\phi d^2 x d^2x',
\label{e:bhel}
\end{equation}
where $\delta \phi$ is the angle between the two field lines $\mathbf{x}$ and $\mathbf{x}'$ as they wind about each other. The following definition 
\begin{equation}
\ell=\frac{1}{\pi}\left | \int_0^L\frac{\dif \phi_{12}}{\dif z}\dif z \right|,
\label{e:w}
\end{equation}
called the linking number, is shown to be a topological invariant \citep{1986GApFD..34..265B} and is used in calculating the total linkage (eg. \citet{2014A&A...564A.131Y}). While equations \eqref{ecross} and \eqref{e:crs} for the crossing number and the winding number involve evaluating the integral of the absolute value, equation \eqref{e:w} for the linking number involves the absolute value of the integral. The reason for the subtle difference is that the linking number integral in equation \eqref{e:w} is the total linkage (an invariant) between the two field lines obtained by integrating over $z$, whereas the crossing number is an energy proxy representing the sum of the fluctuating twists (by taking the absolute value) along $z$. The distribution of linking numbers for the cases of $n=1,3$ and $m=0-3$ are shown in Figure \ref{f:link}.
\begin{figure}[h!]
  \centering
    \begin{subfigure}[]{0.45\textwidth}
    \centering
    \includegraphics[width=1\linewidth]{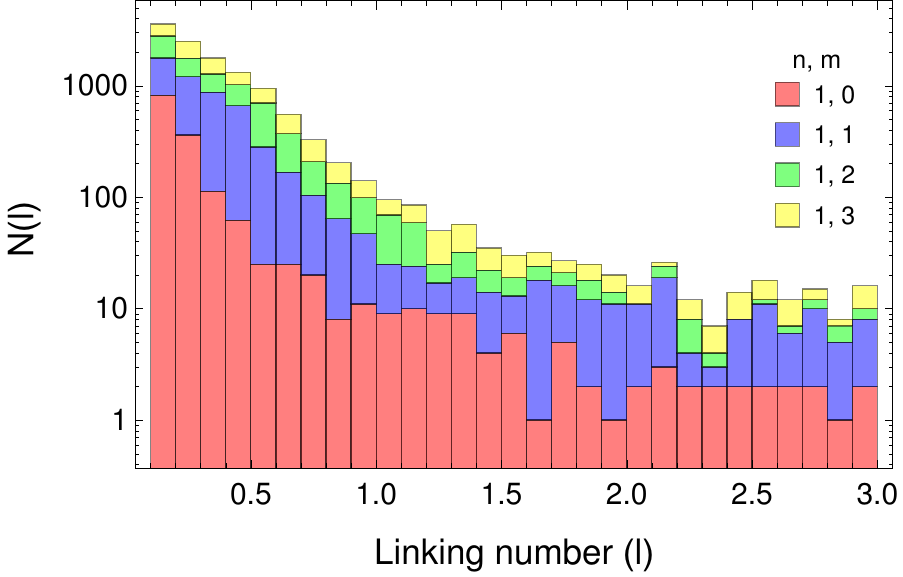}
    \caption{}
  \end{subfigure}
\quad
  \begin{subfigure}[]{0.45\textwidth}
    \centering
    \includegraphics[width=1\linewidth]{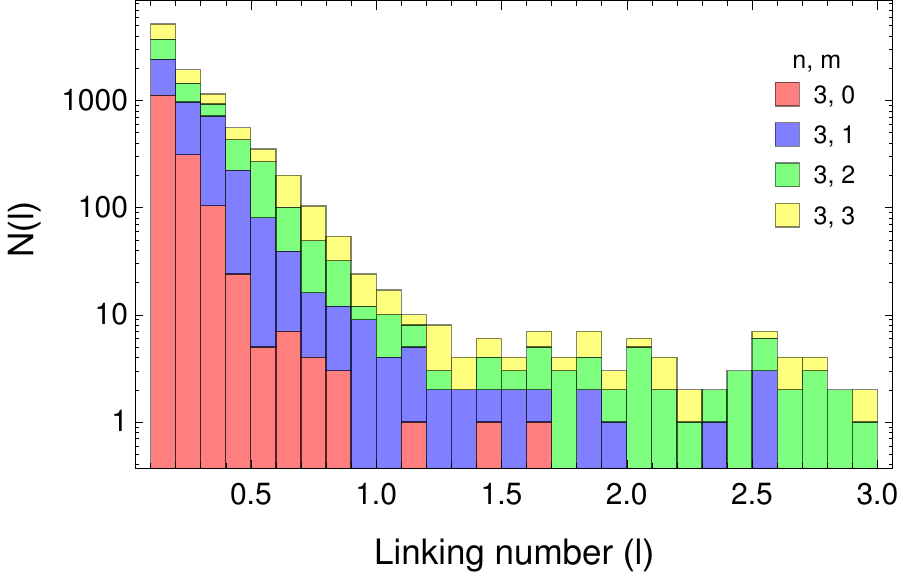}
    \caption{}
  \end{subfigure}
      \caption{Stacked histograms depicting the distribution of number of pairs $N$ for a given values of linking number $l$ calculated for $n =1$, $m = 0-3$ and $n=3$, $m=0-3$ modes.}
  \label{f:link}
\end{figure}
We use the linking number in the following estimate of the relative helicity via linkage in the volume by using the Cauchy-Schwarz inequality:
\begin{align}
{\dif L_r \over \dif z}&= (2\pi)^{-1}\int \int B_z(\mathbf{x})B_z(\mathbf{x}') {\dif \phi \over \dif z}d^2 x d^2x' \\ \nonumber
& \leq B_0^2 \pi^2 R^4 <1+2 \epsilon_z>^{1/2} {1 \over 2 \pi} \left (\dif \phi \over \dif z \right)_{rms} \\ \nonumber
L_r & \leq E  <1+ 2\epsilon_z>^{1/2} {4 \pi^2 R^2 \over L} \ell_{rms}, 
\label{e:bhel2}
\end{align}
where $\ell_{rms}$ is the RMS linking number and $E = B_0^2 \pi R^2 L/ (8 \pi)$ is the energy of the mean magnetic field in $z$ direction. From equation (\ref{Etbound}), we obtain the condition 
\begin{equation}
E E_t \geq { W^2 \over 64 g^2 R^2}, 
\end{equation} 
that leads to another bound
\begin{align}
L_r & \leq 8 g R \sqrt{E_t E},
\end{align}
which follows from the fact that $W > L_r$; this can be seen from equations \eqref{e:w} and  \eqref{e:crs}. These bounds can be used for making
various estimates for given configurations and verifying them for different topologies. The contour plots for the global winding number and relative helicity for the NLFFF solutions arising from crossing between the field lines are shown in Figure \ref{conp}. We find that both the global winding number and relative helicity for the NLFFF solutions increases for higher values of $n$ and $m$ has a stronger positive dependence on $n$ than the parameter $m$.

\begin{figure}[h]
  \centering
  \begin{subfigure}[]{0.45\textwidth}
    \centering
        \includegraphics[width=1\linewidth]{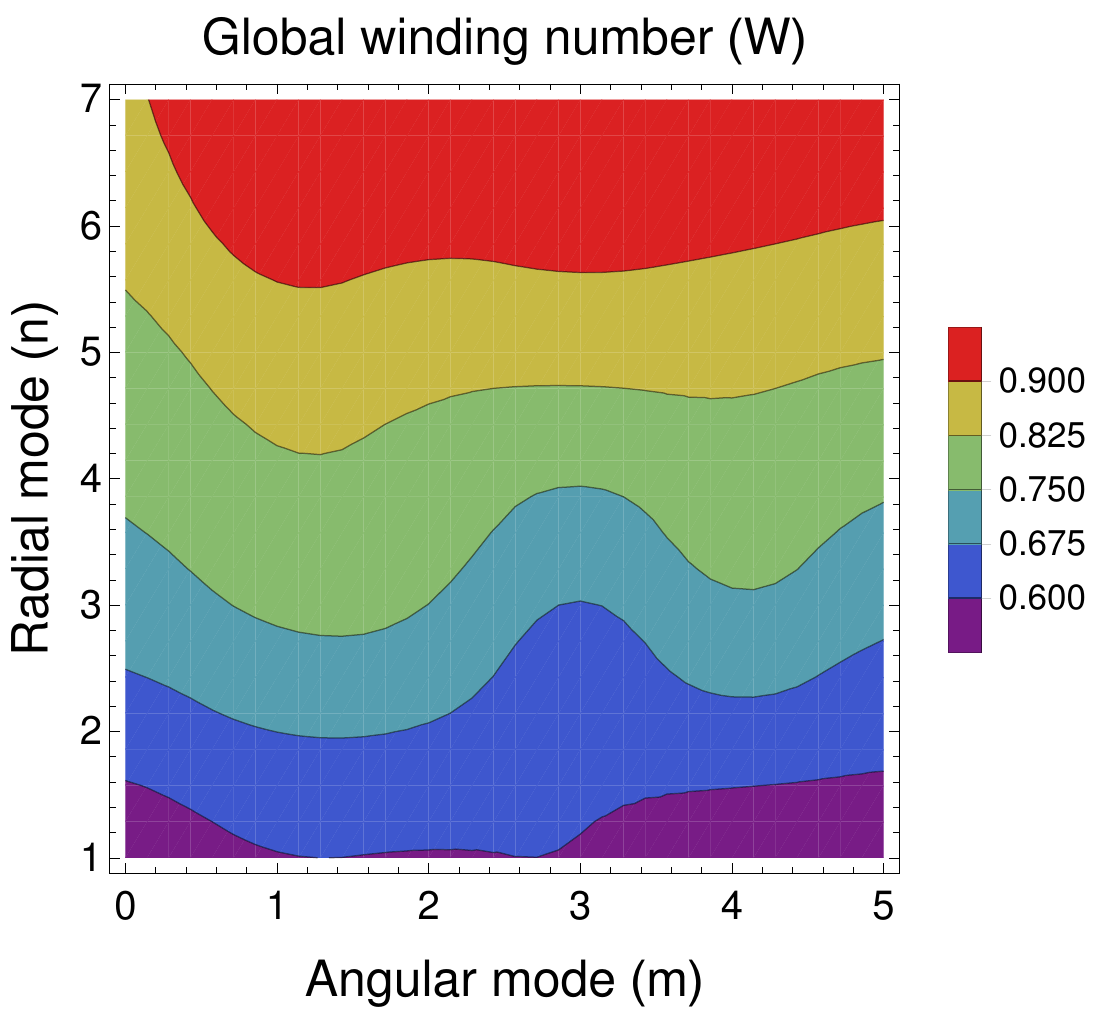}
    \caption{}
  \end{subfigure}
\quad
  \begin{subfigure}[]{0.45\textwidth}
    \centering
    \includegraphics[width=1\linewidth]{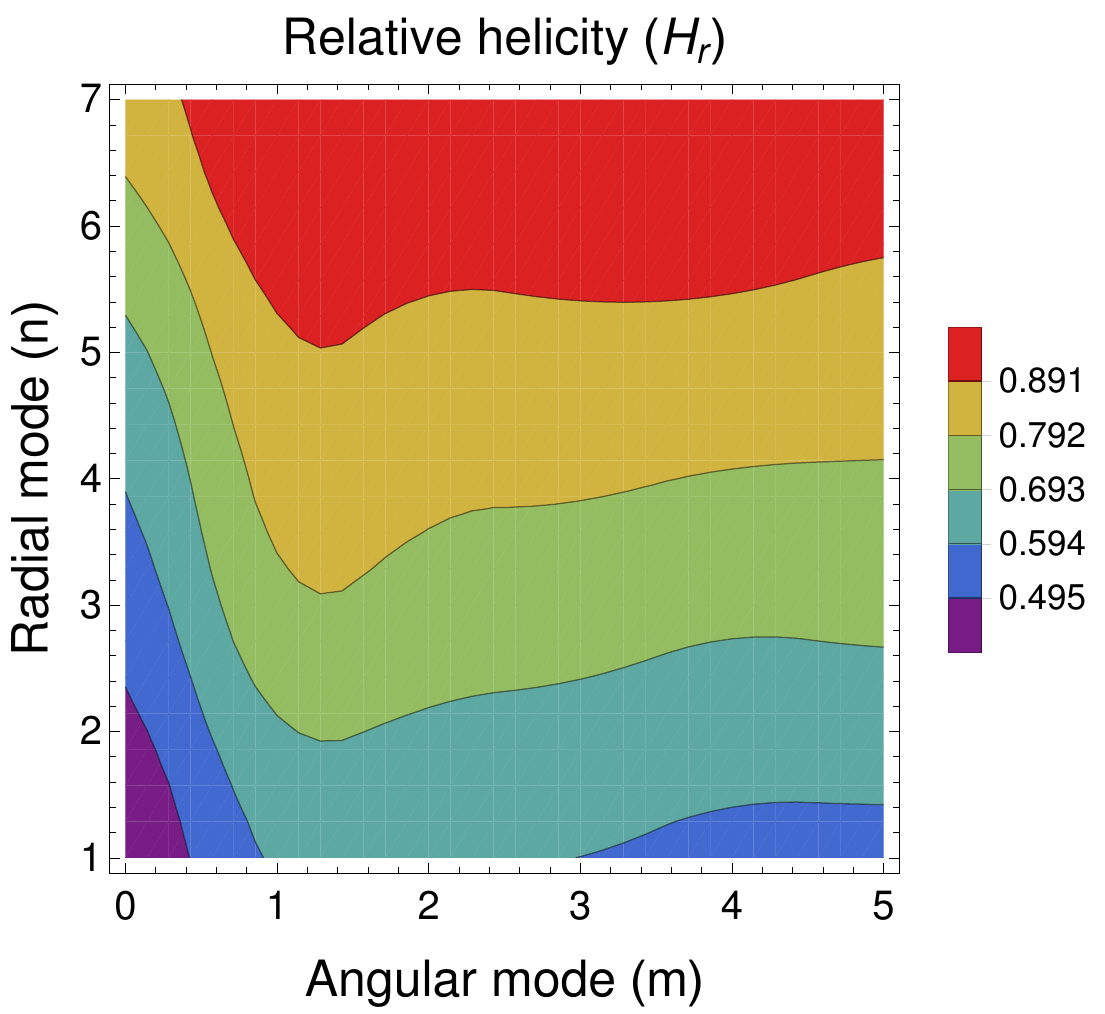}
    \caption{}
  \end{subfigure}
  \caption{Contour plot of (a) global winding number and (b) relative helicity arising due to braiding of magnetic field lines for NLFFF solutions corresponding to $n=1-7$ and $m=0-5$. The contour values have been normalized as we are only interested in the relative strength between the modes.}
  \label{conp}
\end{figure}

\section{Summary and conclusions}
\label{conclude}
The topological complexity of the nonlinear force-free magnetic fields that are applicable to the solar corona are quantified by calculating their crossing, winding and linking number distributions. For this purpose, we present a new formula for the winding number, which incorporates the winding of the magnetic field lines about each other. This is very useful for  cases where the analytical expression for the magnetic field (eg. those presented in \citet{1990ApJ...352..343L} and PMR14) is known. The utility of the winding number formula is first demonstrated for the pair of helices having constant and quadratic dependence on $z$ and also for  NLFFF solutions. The formulae are found to be different due to the effect of viewing angle and a varying vertical field.  We also calculate linking numbers, which are useful in estimating the contribution of magnetic braiding towards the total free energy and relative helicity of the field.  We have presented new analytical bounds for the free energy and relative helicity for the field configurations which are expressed in terms of the linking number that can be verified for different topologies. In future, we plan to explore different models of braiding and reconnection to explain the power-law indices observed in the crossing number and flare-energy distributions. We also plan to  estimate the total energy released in the corona from the braided structures in active regions, and study their significance in the overall context of the energy budget for the active Sun.

We thank the referees for a careful reading of the manuscript and their very useful comments.
\appendix

\section{Numerical estimates of $\kappa$ and $\ol{f^2}$} 

Starting with the integral ${\cal I}_1$ given by
\begin{equation}
R f(a_1)=\mathcal{I}_1 = \int  { \left|\left( \hat{\mbf{t}}_1  \cdot \hat{\boldsymbol{\phi}}_{12}\right ) \right| \over   r_{12}} \dif^2 x_2, 
\end{equation}
we maximize it by choosing $\hat{t}_1(\hat{x}_1)$ in the direction $ \hat{z} \times \hat{x}_1$ so that this produces maximum twist thereby resulting in the following simplification 
\begin{align}
\hat{\mbf{t}}_1  \cdot \hat{\boldsymbol{\phi}}_{12}& =(\hat{z} \times \hat{x}_1) \cdot (\hat{z} \times \hat{r}_{12})=  (\hat{z} \cdot \hat{z})  (\hat{x}_1 \cdot \hat{r}_{12})- (\hat{z} \cdot \hat{r}_{12})  (\hat{x}_1 \cdot \hat{z}) \nonumber \\
&= \hat{x}_1 \cdot \hat{r}_{12}= \frac{(x_2 \cos{\phi}-x_1)}{r_{12}} \nonumber \\
|\hat{\mbf{t}}_1  \cdot \hat{\mbf{\phi}}_{12}| &= \left | \frac{(s \cos{\phi} -1) x_1}{r_{12}} \right | 
\end{align}
where $s\equiv x_2/x_1$ and $\cos{\phi} = \hat{x}_1 \cdot \hat{x}_2$.  We can
now write
\begin{align}
\mathcal{I}_1 &= \int_0^{2 \pi} \dif \phi \int_0^R \frac{|s \cos{\phi} -1| x_1 x_2}{ (x_1^2 +x_2^2 - 2x_1 x_2 \cos{\phi})} \dif x_2 \nonumber \\
 &=  R a_1\int_0^{2 \pi} \dif \phi \int_0^{1/a_1} \frac{|s \cos{\phi} -1| s}{ (1 +s^2 - 2 s \cos{\phi})} \dif s \nonumber \\
f(a_1)&= a_1\int_0^{2 \pi} \dif \phi \int_0^{1/a_1} \frac{|s \cos{\phi} -1| s}{ (1 +s^2 - 2 s \cos{\phi})} \dif s \nonumber \\
\end{align}
where we have used $a_1=x_1/R$. The mean value of $\ol{f^2}$ is given by
\begin{equation}
\ol{f^2}= \int_0^1 2 f^2(a_1) a_1 \dif a_1,
\end{equation}
whose numerical estimate is $\ol{f^2} \simeq 13.14$. The integral for $\kappa$ is given by
\begin{align}
\kappa &= \int_0^{2 \pi} \dif \phi  \int_0^R \dif x_1 \int_0^R \frac{ x_1 x_2}{ (x_1^2 +x_2^2 - 2x_1 x_2 \cos{\phi})} \dif x_2 \nonumber \\
&= \int_0^1 a_1 \dif a_1 \int_0^{2 \pi} \dif \phi \int_0^{1/a_1} \frac{s}{ (1 +s^2 - 2 s \cos{\phi})} \dif s, \nonumber \\
\end{align}
which works out be $\kappa =2 \pi^2 \ln{2}=13.68$.
\bibliography{braids}

\begin{thebibliography}{32}
\expandafter\ifx\csname natexlab\endcsname\relax\def\natexlab#1{#1}\fi
\providecommand{\url}[1]{\texttt{#1}}
\providecommand{\href}[2]{#2}
\providecommand{\path}[1]{#1}
\providecommand{\DOIprefix}{doi:}
\providecommand{\ArXivprefix}{arXiv:}
\providecommand{\URLprefix}{URL: }
\providecommand{\Pubmedprefix}{pmid:}
\providecommand{\doi}[1]{\href{http://dx.doi.org/#1}{\path{#1}}}
\providecommand{\Pubmed}[1]{\href{pmid:#1}{\path{#1}}}
\providecommand{\bibinfo}[2]{#2}
\ifx\xfnm\relax \def\xfnm[#1]{\unskip,\space#1}\fi
\bibitem[{Aschwanden(2004)}]{2004psci.book.....A}
\bibinfo{author}{Aschwanden, M.J.}, \bibinfo{year}{2004}.
\newblock \bibinfo{title}{Physics of the Solar Corona. An Introduction}.
\newblock \bibinfo{edition}{second} ed., \bibinfo{publisher}{Praxis Publishing
  Ltd}, \bibinfo{address}{Chichester, UK}.
\bibitem[{{Berger}(1986)}]{1986GApFD..34..265B}
\bibinfo{author}{{Berger}, M.A.}, \bibinfo{year}{1986}.
\newblock \bibinfo{title}{{Topological invariants of field lines rooted to
  planes.}}
\newblock \bibinfo{journal}{Geophysical and Astrophysical Fluid Dynamics}
  \bibinfo{volume}{34}, \bibinfo{pages}{265--281}.
\bibitem[{{Berger}(1993)}]{1993PhRvL..70..705B}
\bibinfo{author}{{Berger}, M.A.}, \bibinfo{year}{1993}.
\newblock \bibinfo{title}{{Energy-crossing number relations for braided
  magnetic fields}}.
\newblock \bibinfo{journal}{Physical Review Letters} \bibinfo{volume}{70},
  \bibinfo{pages}{705--708}.
\newblock \DOIprefix\doi{10.1103/PhysRevLett.70.705}.
\bibitem[{{Berger} and {Asgari-Targhi}(2009)}]{2009ApJ...705..347B}
\bibinfo{author}{{Berger}, M.A.}, \bibinfo{author}{{Asgari-Targhi}, M.},
  \bibinfo{year}{2009}.
\newblock \bibinfo{title}{{Self-organized Braiding and the Structure of Coronal
  Loops}}.
\newblock \bibinfo{journal}{\apj} \bibinfo{volume}{705},
  \bibinfo{pages}{347--355}.
\newblock \DOIprefix\doi{10.1088/0004-637X/705/1/347}.
\bibitem[{{Berger} et~al.(2015){Berger}, {Asgari-Targhi} and
  {Deluca}}]{2015JPlPh..81d3904B}
\bibinfo{author}{{Berger}, M.A.}, \bibinfo{author}{{Asgari-Targhi}, M.},
  \bibinfo{author}{{Deluca}, E.E.}, \bibinfo{year}{2015}.
\newblock \bibinfo{title}{{Self-organized braiding in solar coronal loops}}.
\newblock \bibinfo{journal}{Journal of Plasma Physics} \bibinfo{volume}{81},
  \bibinfo{pages}{395810404}.
\newblock \DOIprefix\doi{10.1017/S0022377815000483}.
\bibitem[{{Berger} and {Field}(1984)}]{1984JFM...147..133B}
\bibinfo{author}{{Berger}, M.A.}, \bibinfo{author}{{Field}, G.B.},
  \bibinfo{year}{1984}.
\newblock \bibinfo{title}{{The topological properties of magnetic helicity}}.
\newblock \bibinfo{journal}{Journal of Fluid Mechanics} \bibinfo{volume}{147},
  \bibinfo{pages}{133--148}.
\newblock \DOIprefix\doi{10.1017/S0022112084002019}.
\bibitem[{Calugareanu(1959)}]{calugareanu1959integrale}
\bibinfo{author}{Calugareanu, G.}, \bibinfo{year}{1959}.
\newblock \bibinfo{title}{L’int{\'e}grale de gauss et l’analyse des
  n{\oe}uds tridimensionnels ({T}he gaussian integral and analysis of
  three-dimensional knots).}
\newblock \bibinfo{journal}{Rev. Math. pures appl} \bibinfo{volume}{4},
  \bibinfo{pages}{5--20}.
\bibitem[{{Edl{\'e}n}(1943)}]{1943ZA.....22...30E}
\bibinfo{author}{{Edl{\'e}n}, B.}, \bibinfo{year}{1943}.
\newblock \bibinfo{title}{{Die Deutung der Emissionslinien im Spektrum der
  Sonnenkorona. Mit 6 Abbildungen (The interpretation of the emission lines in
  the spectrum of the solar corona).}}
\newblock \bibinfo{journal}{\zap} \bibinfo{volume}{22},
  \bibinfo{pages}{30--64}.
\bibitem[{Freedman and He(1991)}]{freedman1991divergence}
\bibinfo{author}{Freedman, M.H.}, \bibinfo{author}{He, Z.X.},
  \bibinfo{year}{1991}.
\newblock \bibinfo{title}{Divergence-free fields: energy and asymptotic
  crossing number}.
\newblock \bibinfo{journal}{Annals of Mathematics} \bibinfo{volume}{134},
  \bibinfo{pages}{189--229}.
\bibitem[{{Fuller}(1971)}]{1971PNAS...68..815B}
\bibinfo{author}{{Fuller}, F.B.}, \bibinfo{year}{1971}.
\newblock \bibinfo{title}{{The Writhing Number of a Space Curve}}.
\newblock \bibinfo{journal}{Proceedings of the National Academy of Science}
  \bibinfo{volume}{68}, \bibinfo{pages}{815--819}.
\bibitem[{Golub and Pasachoff(2010)}]{2009soco.book.....G}
\bibinfo{author}{Golub, L.}, \bibinfo{author}{Pasachoff, J.M.},
  \bibinfo{year}{2010}.
\newblock \bibinfo{title}{The Solar Corona}.
\newblock \bibinfo{edition}{second} ed., \bibinfo{publisher}{Cambridge
  University Press}, \bibinfo{address}{Cambridge, UK}.
\bibitem[{{Grotrian}(1934)}]{1934ZA......8..124G}
\bibinfo{author}{{Grotrian}, W.}, \bibinfo{year}{1934}.
\newblock \bibinfo{title}{{{\"U}ber das Fraunhofersche Spektrum der
  Sonnenkorona. Mit 10 Abbildungen (About the Fraunhofer spectrum of the solar
  corona).}}
\newblock \bibinfo{journal}{\zap} \bibinfo{volume}{8},
  \bibinfo{pages}{124--146}.
\bibitem[{{Ionson}(1985)}]{1985SoPh..100..289I}
\bibinfo{author}{{Ionson}, J.A.}, \bibinfo{year}{1985}.
\newblock \bibinfo{title}{{The heating of coronae}}.
\newblock \bibinfo{journal}{\solphys} \bibinfo{volume}{100},
  \bibinfo{pages}{289--308}.
\newblock \DOIprefix\doi{10.1007/BF00158433}.
\bibitem[{{Klimchuk}(2006)}]{2006SoPh..234...41K}
\bibinfo{author}{{Klimchuk}, J.A.}, \bibinfo{year}{2006}.
\newblock \bibinfo{title}{{On Solving the Coronal Heating Problem}}.
\newblock \bibinfo{journal}{\solphys} \bibinfo{volume}{234},
  \bibinfo{pages}{41--77}.
\newblock \DOIprefix\doi{10.1007/s11207-006-0055-z},
  \href{http://arxiv.org/abs/astro-ph/0511841}{\tt arXiv:astro-ph/0511841}.
\bibitem[{{Kumar} et~al.(2016){Kumar}, {Bhattacharyya}, {Joshi} and
  {Smolarkiewicz}}]{2016ApJ...830...80K}
\bibinfo{author}{{Kumar}, S.}, \bibinfo{author}{{Bhattacharyya}, R.},
  \bibinfo{author}{{Joshi}, B.}, \bibinfo{author}{{Smolarkiewicz}, P.K.},
  \bibinfo{year}{2016}.
\newblock \bibinfo{title}{{On the Role of Repetitive Magnetic Reconnections in
  Evolution of Magnetic Flux Ropes in Solar Corona}}.
\newblock \bibinfo{journal}{\apj} \bibinfo{volume}{830}, \bibinfo{pages}{80}.
\newblock \DOIprefix\doi{10.3847/0004-637X/830/2/80},
  \href{http://arxiv.org/abs/1609.08260}{\tt arXiv:1609.08260}.
\bibitem[{{Low} and {Lou}(1990)}]{1990ApJ...352..343L}
\bibinfo{author}{{Low}, B.C.}, \bibinfo{author}{{Lou}, Y.Q.},
  \bibinfo{year}{1990}.
\newblock \bibinfo{title}{{Modeling solar force-free magnetic fields}}.
\newblock \bibinfo{journal}{\apj} \bibinfo{volume}{352},
  \bibinfo{pages}{343--352}.
\newblock \DOIprefix\doi{10.1086/168541}.
\bibitem[{{Mandrini} et~al.(2000){Mandrini}, {D{\'e}moulin} and
  {Klimchuk}}]{2000ApJ...530..999M}
\bibinfo{author}{{Mandrini}, C.H.}, \bibinfo{author}{{D{\'e}moulin}, P.},
  \bibinfo{author}{{Klimchuk}, J.A.}, \bibinfo{year}{2000}.
\newblock \bibinfo{title}{{Magnetic Field and Plasma Scaling Laws: Their
  Implications for Coronal Heating Models}}.
\newblock \bibinfo{journal}{\apj} \bibinfo{volume}{530},
  \bibinfo{pages}{999--1015}.
\newblock \DOIprefix\doi{10.1086/308398}.
\bibitem[{{Milano} et~al.(1997){Milano}, {G{\'o}mez} and
  {Martens}}]{1997ApJ...490..442M}
\bibinfo{author}{{Milano}, L.J.}, \bibinfo{author}{{G{\'o}mez}, D.O.},
  \bibinfo{author}{{Martens}, P.C.H.}, \bibinfo{year}{1997}.
\newblock \bibinfo{title}{{Solar Coronal Heating: AC versus DC}}.
\newblock \bibinfo{journal}{\apj} \bibinfo{volume}{490},
  \bibinfo{pages}{442--451}.
\bibitem[{{Moffatt} and {Ricca}(1992)}]{1992RSPSA.439..411M}
\bibinfo{author}{{Moffatt}, H.K.}, \bibinfo{author}{{Ricca}, R.L.},
  \bibinfo{year}{1992}.
\newblock \bibinfo{title}{{Helicity and the Calugareanu Invariant}}.
\newblock \bibinfo{journal}{Proceedings of the Royal Society of London Series
  A} \bibinfo{volume}{439}, \bibinfo{pages}{411--429}.
\newblock \DOIprefix\doi{10.1098/rspa.1992.0159}.
\bibitem[{{Parker}(1972)}]{1972ApJ...174..499P}
\bibinfo{author}{{Parker}, E.N.}, \bibinfo{year}{1972}.
\newblock \bibinfo{title}{{Topological Dissipation and the Small-Scale Fields
  in Turbulent Gases}}.
\newblock \bibinfo{journal}{\apj} \bibinfo{volume}{174},
  \bibinfo{pages}{499--510}.
\newblock \DOIprefix\doi{10.1086/151512}.
\bibitem[{Parker(1979)}]{1979cmft.book.....P}
\bibinfo{author}{Parker, E.N.}, \bibinfo{year}{1979}.
\newblock \bibinfo{title}{Cosmical magnetic fields: Their origin and their
  activity}.
\newblock \bibinfo{publisher}{Clarendon Press}, \bibinfo{address}{Oxford}.
\bibitem[{{Parker}(1983)}]{1983ApJ...264..642P}
\bibinfo{author}{{Parker}, E.N.}, \bibinfo{year}{1983}.
\newblock \bibinfo{title}{{Magnetic Neutral Sheets in Evolving Fields - Part
  Two - Formation of the Solar Corona}}.
\newblock \bibinfo{journal}{\apj} \bibinfo{volume}{264},
  \bibinfo{pages}{642--647}.
\newblock \DOIprefix\doi{10.1086/160637}.
\bibitem[{{Parker}(1988)}]{1988ApJ...330..474P}
\bibinfo{author}{{Parker}, E.N.}, \bibinfo{year}{1988}.
\newblock \bibinfo{title}{{Nanoflares and the solar X-ray corona}}.
\newblock \bibinfo{journal}{\apj} \bibinfo{volume}{330},
  \bibinfo{pages}{474--479}.
\newblock \DOIprefix\doi{10.1086/166485}.
\bibitem[{{Parker}(1994)}]{1994ISAA....1.....P}
\bibinfo{author}{{Parker}, E.N.}, \bibinfo{year}{1994}.
\newblock \bibinfo{title}{{Spontaneous current sheets in magnetic fields : with
  applications to stellar x-rays}}.
\newblock \bibinfo{journal}{Spontaneous current sheets in magnetic fields :
  with applications to stellar x-rays.~ International Series in Astronomy and
  Astrophysics, Vol.~1.~ New York : Oxford University Press, 1994.}
  \bibinfo{volume}{1}.
\bibitem[{Prasad and Mangalam(2013)}]{2013ASInC..10...51P}
\bibinfo{author}{Prasad, A.}, \bibinfo{author}{Mangalam, A.},
  \bibinfo{year}{2013}.
\newblock \bibinfo{title}{Models of force-free spheres and applications to
  solar active regions}, in: \bibinfo{editor}{{Gopalswamy}, N.},
  \bibinfo{editor}{{Hasan}, S.S.}, \bibinfo{editor}{{Rao}, P.B.},
  \bibinfo{editor}{{Subramanian}, P.} (Eds.), \bibinfo{booktitle}{ASI
  Conference Series, Vol. 10}, pp. \bibinfo{pages}{51--57}.
\bibitem[{{Prasad} et~al.(2014){Prasad}, {Mangalam} and
  {Ravindra}}]{2014ApJ...786...81P}
\bibinfo{author}{{Prasad}, A.}, \bibinfo{author}{{Mangalam}, A.},
  \bibinfo{author}{{Ravindra}, B.}, \bibinfo{year}{2014}.
\newblock \bibinfo{title}{{Separable Solutions of Force-Free Spheres and
  Applications to Solar Active Regions}}.
\newblock \bibinfo{journal}{\apj} \bibinfo{volume}{786}, \bibinfo{pages}{81}.
\newblock \DOIprefix\doi{10.1088/0004-637X/786/2/81},
  \href{http://arxiv.org/abs/1404.0910}{\tt arXiv:1404.0910}.
\bibitem[{Schrijver and Zwaan(2000)}]{2000ssma.book.....S}
\bibinfo{author}{Schrijver, C.J.}, \bibinfo{author}{Zwaan, C.},
  \bibinfo{year}{2000}.
\newblock \bibinfo{title}{Solar and Stellar Magnetic Activity}.
\newblock Cambridge astrophysics series ; 34, \bibinfo{publisher}{Cambridge
  University Press}, \bibinfo{address}{New York}.
\bibitem[{{van Ballegooijen} et~al.(2014){van Ballegooijen}, {Asgari-Targhi}
  and {Berger}}]{2014ApJ...787...87V}
\bibinfo{author}{{van Ballegooijen}, A.A.}, \bibinfo{author}{{Asgari-Targhi},
  M.}, \bibinfo{author}{{Berger}, M.A.}, \bibinfo{year}{2014}.
\newblock \bibinfo{title}{{On the Relationship Between Photospheric Footpoint
  Motions and Coronal Heating in Solar Active Regions}}.
\newblock \bibinfo{journal}{\apj} \bibinfo{volume}{787}, \bibinfo{pages}{87}.
\newblock \DOIprefix\doi{10.1088/0004-637X/787/1/87}.
\bibitem[{{Wiegelmann} and {Sakurai}(2012)}]{2012LRSP....9....5W}
\bibinfo{author}{{Wiegelmann}, T.}, \bibinfo{author}{{Sakurai}, T.},
  \bibinfo{year}{2012}.
\newblock \bibinfo{title}{{Solar Force-free Magnetic Fields}}.
\newblock \bibinfo{journal}{Living Reviews in Solar Physics}
  \bibinfo{volume}{9}, \bibinfo{pages}{5}.
\newblock \DOIprefix\doi{10.12942/lrsp-2012-5},
  \href{http://arxiv.org/abs/1208.4693}{\tt arXiv:1208.4693}.
\bibitem[{{Wilmot-Smith} et~al.(2009){Wilmot-Smith}, {Hornig} and
  {Pontin}}]{2009ApJ...696.1339W}
\bibinfo{author}{{Wilmot-Smith}, A.L.}, \bibinfo{author}{{Hornig}, G.},
  \bibinfo{author}{{Pontin}, D.I.}, \bibinfo{year}{2009}.
\newblock \bibinfo{title}{{Magnetic Braiding and Parallel Electric Fields}}.
\newblock \bibinfo{journal}{\apj} \bibinfo{volume}{696},
  \bibinfo{pages}{1339--1347}.
\newblock \DOIprefix\doi{10.1088/0004-637X/696/2/1339},
  \href{http://arxiv.org/abs/0810.1415}{\tt arXiv:0810.1415}.
\bibitem[{{Withbroe} and {Noyes}(1977)}]{1977ARA&A..15..363W}
\bibinfo{author}{{Withbroe}, G.L.}, \bibinfo{author}{{Noyes}, R.W.},
  \bibinfo{year}{1977}.
\newblock \bibinfo{title}{{Mass and energy flow in the solar chromosphere and
  corona}}.
\newblock \bibinfo{journal}{\araa} \bibinfo{volume}{15},
  \bibinfo{pages}{363--387}.
\newblock \DOIprefix\doi{10.1146/annurev.aa.15.090177.002051}.
\bibitem[{{Yeates} et~al.(2014){Yeates}, {Bianchi}, {Welsch} and
  {Bushby}}]{2014A&A...564A.131Y}
\bibinfo{author}{{Yeates}, A.R.}, \bibinfo{author}{{Bianchi}, F.},
  \bibinfo{author}{{Welsch}, B.T.}, \bibinfo{author}{{Bushby}, P.J.},
  \bibinfo{year}{2014}.
\newblock \bibinfo{title}{{The coronal energy input from magnetic braiding}}.
\newblock \bibinfo{journal}{\aap} \bibinfo{volume}{564}, \bibinfo{pages}{A131}.
\newblock \DOIprefix\doi{10.1051/0004-6361/201323276},
  \href{http://arxiv.org/abs/1403.4396}{\tt arXiv:1403.4396}.

\end{thebibliography}

\end{document}